# Dynamical electronic correlation and chiral magnetism in van der Waals magnet Fe$_4$GeTe$_2$

Md. Nur Hasan[1], Nastaran Salehi[1], Felix Sorgenfrei[1], Anna Delin[3,4,5], Igor Di Marco[1,6], Anders Bergman[1], Manuel Pereiro[1], Patrik Thunström[1], Olle Eriksson[1,2] and Debjani Karmakar[1,7,8,*]

[1] *Department of Physics and Astronomy,*
*Uppsala University, Box 516, SE-751 20, Uppsala, Sweden*

[2] *WISE, Wallenberg Initiative Materials Science,*
*Uppsala University, Box 516, SE-751 20, Uppsala, Sweden*

[3] *Department of Applied Physics*
*KTH Royal Institute of Technology*
*SE-106 91 Stockholm, Sweden*

[4] *Swedish e-Science Research Center (SeRC),*
*KTH Royal Institute of Technology,*
*SE-10044 Stockholm, Sweden*

[5] *WISE, Wallenberg Initiative Materials Science,*
*KTH Royal Institute of Technology,*
*SE-10044 Stockholm, Sweden*

[6] *Institute of Physics, Faculty of Physics, Astronomy & Informatics,*
*Nicolaus Copernicus University, Grudziadzka 5, 87-100, Torun, Poland*

[7] *Technical Physics Division, Bhabha Atomic Research Centre, Mumbai 400085, India*

[8] *Homi Bhabha National Institute,*
*Anushaktinagar, Mumbai 400094, India.*

* Corresponding Author:
    Debjani Karmakar
    Email: debjani.karmakar@physics.uu.se

**Abstract**

Among the quasi-2D van der Waals magnetic systems, $Fe_4GeTe_2$ imprints a profound impact due to its near-room temperature ferromagnetic behaviour and the complex magnetothermal phase diagram exhibiting multiple phase transformations, as observed from magnetization and magnetotransport measurements. A complete analysis of these phase transformations in the light of electronic correlation and its impact on the underlying magnetic interactions remain unattended in the existing literature. Using first-principles methodologies, incorporating the dynamical nature of electron correlation, we have analysed the interplay of the direction of magnetization in the easy-plane and easy-axis manner with the underlying crystal symmetry, which reveals the opening of a pseudogap feature beyond the spin-reorientation transition (SRT) temperature. The impact of dynamical correlation on the calculated magnetic circular dichroism and x-ray absorption spectrum of the L-edge of the Fe atoms compared well with the existing experimental observations. The calculated intersite Heisenberg exchange interactions display a complicated nature, depending upon the pairwise interactions among the two inequivalent Fe sites, indicating a RKKY-like behaviour of the magnetic interactions. We noted the existence of significant anisotropic and antisymmetric exchanges interactions, resulting into a chirality in the magnetic behaviour of the system. Subsequent investigation of the dynamical aspects of magnetism in $Fe_4GeTe_2$ and the respective magnetothermal phase diagram reveal that the dynamical nature of spins and the decoupling of the magnetic properties for both sites of Fe is crucial to explain all the experimentally observed phase transformations.

# I. INTRODUCTION

The sustained quest for the new 2D van der Waals (vdW) systems has approached a new horizon after obtaining access to the optimal control of spin and orbital magnetic moments and their respective coupling. In 2D magnetic systems, the superiority of the transport and opto-electronic properties are boosted by the magnetic and topological control to generate altogether new functionalities, hitherto unattainable in their 3D counterparts [1-5]. Interestingly, a fair share of such 2D materials [6,7] were predicted from first-principles theory [8]. For such systems, the magnetocrystalline (MA) and shape anisotropies (SA) can be utilized as an escape route to the restriction imposed by the Mermin-Wagner theorem regarding the absence of a finite-temperature long-range order in the 2D limit [9]. Following this pathway, systems like metallic trichalcogenides [10-12], chromium-based trihalides [13-15] and $Cr_2Ge_2Te_6$ [16] were seen to retain a 2D-ferromagnetic order, albeit with a lower Curie temperature ($T_c$). The simultaneous achievement of a lower dimensionality and a higher $T_c$ in a material can promote its utilization in spin-channel, spin-source or memory devices [17,18].

In conventional 2D magnetic systems, the magnetic layer is encapsulated within the non-magnetic layers. Therefore, however strong the in-plane magnetic interaction be, achieving a high $T_c$ is difficult due to the quenched interlayer magnetic coupling across the structural gap linked by weak vdW force. In recent times, $Fe_nGeTe_2$ ($n$ = 3,4,5) (FnGT) were identified with the potential to generate a room-temperature ferromagnetic (FM) order. Here, an intralayer 3D-like magnetic network was created after incorporating Fe within a Ge-layer [19-22]. A few such magnetic layers are encapsulated within the Te-ligand layers. Such a diligent methodology to enhance the $T_c$ is, however, associated with a trade-off of driving the system into a quasi-2D one, where the common synthesis routes experience nonstoichiometric samples [22-24]. The experimentally obtained $T_c$ for $n$ = 3, 4 and 5 are 220 K, 280 K and 310 K respectively, indicating a monotonous rise of $T_c$ with an increasing Fe content [25-27]. The first member of this series, F3GT, is an itinerant ferromagnet with a large out-of-plane MA [21]. For the third system F5GT with the highest $T_c$, the relation between the stoichiometry and the crystal structure is still an open question [22,28].

The middle member, $Fe_4GeTe_2$ (F4GT), is experimentally known to be the most interesting one, where, the magnetic and magnetothermal measurements revealed a multitude of phase-transformations, indicating an underlying complex interplay of electron correlation, temperature-dependent magnetocrystalline anisotropy, field-induced symmetry breaking, X-ray spectral properties, anisotropic intersite exchange and dynamical magnetism. A complete picture of all these tangled attributes is not yet available in the existing literature, leading to the generation of some important unanswered questions. First, albeit the expected designed order for F4GT is ferromagnetic, the existing experiments indicate a more complex chiral order in this system [26, 32]. A clear knowledge of this complex magnetic behaviour and its implicit or explicit impacts on the experimental magnetic attributes are not yet available. Second, the reason behind its multiple magnetothermal phase transitions and its dependence with the underlying crystal symmetries is not known. Third, the experimentally obtained temperature-dependent crossover of the magnetoresistance from positive to negative values is not yet analysed [31]. Fourth, the anomalous Hall measurements indicate the presence of topological band-crossings near Fermi-energy, which was not clearly observed in the band structure obtained within the simpler formulations of density functional theory (DFT) or DFT + Hubbard U based treatment of electron correlations [29-31]. Fifth, various magnetothermal phase-transitions and its interconnection with the dynamical magnetic ground-states are not yet explored, where an interplay of the temperature-dependent MA with its magnetotransport properties leads to a complex phase diagram [19,26,29-32].

Keeping in mind this defragmented status of the analysis of F4GT from the theoretical as well as from the experimental studies, we present a coherent investigation of electronic properties of F4GT after incorporating the dynamical nature of electron correlation with DFT + Dynamical mean-field theory (DMFT) after incorporating spin-orbit coupling (SOC), followed by an extraction of the intersite isotropic and anisotropic exchange parameters and a subsequent spin-dynamical calculations. The appropriate treatment of electron correlation has enabled us to obtain a clear explanation of field-induced symmetry breaking and also has established the predicted presence of topological band-crossings. On the other hand, the derived spin-dynamical ground states under various thermomagnetic conditions can provide a natural interpretation of some of the experimentally observed results in the light of a microscopic and atomistic analysis. Moreover, these results can also be used as a future reference to interpret future more complex and detailed experiments on F4GT.

The different sections of this work are organized as follows. In the first section, we elaborate on the unattended and unanswered questions related to the magnetic properties of F4GT. Next, we elaborate on the computational methodologies, which are used for a detailed analysis of this system. In the successive section, we describe the structural attributes of this system and the subsequent electronic structure analysis with the help of a combined DFT+DMFT+SOC calculation. In the following section, we have analysed the x-ray spectral properties of this system using the DFT+DMFT combined with multiplet ligand field theory (MLFT). This was complemented by an elaborate computation of exchange parameters by using a DFT+DMFT coupled to the Lichtenstein-Katsnelson-Antropov-Gubanov (LKAG) method and subsequent spin-dynamical calculations. Using these wide range of methodologies, we have investigated the role of dynamical electron-electron correlation on the electronic structure, x-ray spectral properties, intersite magnetic interactions and dynamical magnetic properties, which indicated a possible pathway to elucidate the complex experimental phase-diagram.

## II. F4GT: UNATTENDED AND UNRESOLVED QUERIES

F4GT undergoes two primary phase transitions with varying temperature. The first one is a paramagnetic to ferromagnetic second-order transition at ~ 270 K and the second one is a first-order SRT at ~ 100 K ($T_{SRT}$) [29-32]. Recent magnetoresistance (MR) measurements indicate another electronic transition at ~ 40 K ($T_P$), where the MR changes its behaviour from a quadratic to a linear temperature dependence [31]. The fundamental reason behind these phase transitions and their impact on the underlying magnetic properties is still not transparent from the existing studies.

For F3GT, the interaction of different magnetic local moments with itinerant electrons resulting in a Kondo screening has been verified by inelastic neutron scattering measurements [20]. A similar complete analysis of the long range order is not available for F4GT. A comparison of the x-ray absorption (XAS) and magnetic circular dichroism (XMCD) measurements on both F3GT and F4GT indicates that for F4GT, the orbital occupancy and the associated electronic structure are more covalent and isotropic [33]. In line with this observation, the experimentally obtained MA for F4GT is one-sixth of that of F3GT [33]. As of now, there is no theoretical study of the X-ray spectral properties of F4GT, after taking care of its complicated magnetic nature.

The magnetothermal and magnetotransport experiments of F4GT have many unresolved aspects. The MR measurements indicate three different temperature ($T$) windows, *viz.* 1) $T < T_P$, with a positive MR-value and a $T^2$ dependence, 2) $T_P < T < T_{SRT}$, with a negative MR and mixed $T + T^2$ dependence, and 3) $T > T_{SRT}$, with a linear $T$-dependence [31]. The scattering mechanism in the third window is experimentally concluded to be the electron-phonon coupling [31]. However, the same mechanism for the first two windows is predicted to be an admixture of electron-electron, electron-magnon and magnon-phonon coupling. The actual underlying mechanism behind the temperature dependence of MR and the origin of its crossover to a negative value at $T_\text{p}$ are not yet understood. Thus, a detailed temperature dependent investigation of the static and dynamic magnetic attributes of this system is necessary to address these unanswered questions.

In F4GT, the presence of an AHE is explained by using the intrinsic Karplus-Luttinger mechanism related to the spin-orbit coupling (SOC)-induced finite Berry curvature, indicating a possible presence of topological band-crossings near the Fermi energy ($E_\text{F}$) [30]. F4GT also bears a striking contrast with the conventional vdW ferromagnets, where a decrease in $T_\text{c}$ is observed with lowering thickness [24]. The thickness-dependence of $T_\text{c}$, the high sensitivity of the dependence of MA on the orientation of the easy axis and their respective temperature-dependent nature indicate the possible presence of a complex magnetic texture [24].

In the ultrathin 2D limit, the ferromagnetic $T_\text{c}$ of an N-atom coordinated system in presence of a nearest-neighbour exchange parameter $J$ can be approximated as $T_\text{c} = T_\text{B}/(\ln[3\pi T_\text{B}/4K])$, where $T_\text{B} = 4\pi J/N$ is the ferromagnetic $T_\text{c}$ of the 3D counterpart and $K$ being the MA [34]. Therefore, any finite anisotropy in the 2D system ensures a finite $T_\text{c}$. In general, 2D systems display a thickness dependent directional crossover for MA, transforming from the easy-axis to the easy-plane type beyond a critical thickness [34]. On the contrary, F4GT displays far more complicated behaviour, owing to the temperature-dependent crossover of the MA [29]. In this system, the interdependence of the electron-electron correlation, spin-orbit coupling and the magnetic exchange interactions with this directional crossovers of MA are not yet explored elaborately.

# III. COMPUTATIONAL METHODS USED IN THE ANALYSIS

## A. DFT + Dynamical mean-field theory (DMFT) + SOC calculations

We have used DFT+DMFT+SOC approach to treat the moderately correlated F4GT system. In such calculations, the basic problem is to search for the actual set of "correlated orbitals" $\{|R, \xi >\}$. Here, R and $\xi$ are the Bravais lattice site index and the orbital index of the correlated atom respectively. For correlated systems, the total Hamiltonian of the system can be written in terms of a Hubbard-like Hamiltonian:

$$H = H_{LDA} + \frac{1}{2}\sum_R \sum_{\xi_1\ldots\xi_4} U_{\xi_1\ldots\xi_4} c^\dagger_{R,\xi_1} c^\dagger_{R,\xi_2} c_{R,\xi_3} c_{R,\xi_4}. \tag{1}$$

Here, $H_{LDA}$ is the computed Hamiltonian under local density approximation. For the atomistic orbitals, the local Coulomb repulsion parameter $U_{\xi_1\ldots\xi_4}$ is related to the Slater integrals $F^n$ as:

$$U_{\xi_1\ldots\xi_4} = \sum_{n=0}^{2l} a_n(\xi_1 \ldots \xi_4) F_n \tag{2}$$

$$a_n(\xi_1 \ldots \xi_4) = \frac{4\pi}{2n+1}\sum_{q=-n}^{n}\langle\xi_1|Y_{nq}|\xi_3\rangle\langle\xi_2|Y^*_{nq}|\xi_4\rangle \tag{3}$$

here $\langle\xi_1|Y_{nq}|\xi_3\rangle$ and $\langle\xi_2|Y^*_{nq}|\xi_4\rangle$ are the integrals over the products of three spherical harmonics. The solution of this effective Hubbard model within spectral DFT involves an approximation imposed upon the many body system, where we map the original system into another with lesser degrees of freedom under the constraint that the expectation value of the observables remain intact. In DFT, the main observable is defined as the charge density $\rho(r)$. In spectral DFT, the one electron Green's function is written as:

$$\hat{G}(z) = \left[(z-\mu)\mathbb{I} - \hat{h}_{LDA} - \hat{\Sigma}(z)\right]^{-1} \tag{4}$$

here $z$ is the energy in the complex plane, $\mu$ is the chemical potential, $\hat{h}_{LDA}$ is the unperturbed LDA Hamiltonian with the hopping term. The most important term representing the electronic correlations is the self-energy operator, $\hat{\Sigma}(z)$, which is a many-body representative of the electron interactions. In spectral DFT, the main observable is the local Green's function at the site R:

$$\hat{G}_R(z) = \hat{P}_R \hat{G}(z) \hat{P}_R \tag{5}$$

$\hat{P}_R = \sum_\xi |R,\xi\rangle\langle R,\xi|$ is the projection operator in the correlated subspace $\{|R,\xi>\}$. Similar to the LDA or GGA approximations in DFT, spectral DFT deals with Dynamical mean field theory (DMFT) approximation, with a purely "local" self-energy [35]. This assumption allows us to concentrate only on single site R and the effect of the other sites can be replaced with a self-consistent electronic bath $\mathcal{G}_0^{-1}(R,z)$. It is assumed that the atomic "impurity" site is embedded in a fermionic bath and that system is treated within a multiband Anderson impurity model. Since we have little idea about the exact Hamiltonian, the effective action can be written as:

$$S = -\iint d\tau d\tau' \sum_{\xi_1 \xi_2} c^\dagger_{\xi_1}(\tau') [\mathcal{G}_0^{-1}]_{\xi_1 \xi_2}(\tau-\tau') c_{\xi_2}(\tau) +$$
$$\frac{1}{2}\int d\tau \sum_{\xi_1\ldots\xi_4} c^\dagger_{\xi_1}(\tau) c^\dagger_{\xi_2}(\tau) U_{\xi_1\ldots\xi_4} c_{\xi_4}(\tau) c_{\xi_3}(\tau) \tag{6}$$

here $\tau$ is the imaginary time for the finite temperature many-body system and the lower and upper limit of the integral are 0 and $\frac{1}{KT}$ respectively. Therefore, the dynamics within the problem is now imposed through the action $S$. The self-energy operator can thus be explicitly calculated from the impurity Green's function $\hat{G}_{imp}(z)$ with the help of the inverse Dyson equation after solving the "impurity" problem:

$$\hat{\Sigma}(R,z) = \mathcal{G}_0^{-1}(R,z) - G_{imp}^{-1}(z) \tag{7}$$

This multiband Anderson impurity model is solved with the help of the spin-polarized T-matrix fluctuation-exchange (SPTF) impurity solver [36, 37], which is known to be efficient and reliable for moderately correlated electron systems. After solving the effective impurity problem, one obtains the updated self-energy. The number of particles will also undergo an effective change and thus there will be a respective modification of the chemical potential. In this way, a new single-particle Green's function is obtained and thus one may now derive a new electronic bath $\mathcal{G}_0^{-1}(R,z)$ by using the inverse Dyson equation:

$$\mathcal{G}_0^{-1}(R,z) = G_R^{-1}(z) + \hat{\Sigma}(R,z) . \tag{8}$$

This full iterative process is continued until a convergent self-energy is obtained. Next, we back-project this self-energy onto the DFT Hamiltonian and obtain a convergence of the electronic charge via the full self-consistent cycle to obtain a complete solution of DFT+DMFT. For any correlated system, the imaginary part ($\Im$) of the calculated self-energy are related to the lifetimes($\tau$) of the quasiparticles as:

$$\tau = \frac{\hbar}{\Im(\hat{\Sigma}(R,z))} \tag{9}$$

All of these calculations are done after using the relativistic spin polarized toolkit (RSPt) [38-40], which is based on a full-potential description using linearized muffin-tin orbitals. The *k*-point mesh used for Brillouin zone sampling was 12×12×6 for all these calculations. The self-consistently converged DFT+DMFT+SOC calculations are the starting point of all the following analysis.

**B. Inter-site exchange interactions : DFT + DMFT + LKAG**

We have obtained the inter-site exchange interactions like the Heisenberg and DM interactions after using the Lichtenstein-Katnelson-Antropov-Gubanov (LKAG) formulation [41,42], where we map the concerned electronic system onto a generalized classical Heisenberg model:

$$H = -\sum_{i \neq j} e_i^\alpha \hat{J}_{ij}^{\alpha\beta} e_j^\beta , \quad \alpha, \beta = x, y, z \tag{10}$$

The unit vector $e_i$ is the local spin direction. The full exchange tensor have interactions like Heisenberg exchange, DM interaction and symmetric anisotropic exchange $\hat{\Gamma}$, which can be expressed for the z-component as:

$$D_{ij}^z = \frac{1}{2}\left(J_{ij}^{xy} - J_{ij}^{yx}\right) \tag{11}$$

$$\Gamma_{ij}^z = \frac{1}{2}\left(J_{ij}^{xy} + J_{ij}^{yx}\right) . \tag{12}$$

The x and y components can be calculated from similar expressions. We extract these parameters from a fully self-consistent converged DFT+DMFT calculation, according to the reference [43].

**C. X-ray spectral properties: DFT+DMFT+MLFT**

We have used the relaxed bulk structure as an input for the computation of the X-ray spectroscopic properties. The fully relativistic self-consistent DFT ground state is obtained with the LDA potentials, after incorporating the SOC, as implemented in full potential Linearized Muffin-Tin orbital (FP-LMTO) based RSPt code [38, 44]. The projection to the $3d$ orbitals of Fe was used to compute the projected density of states (DOS), the hybridization function and the spin-orbit coupling parameters. To compute the $L_{2,3}$ edges, the core-hole is created at the Fe-$2p$ levels and the corresponding $2p$-levels are explicitly included in the basis set (valence levels) for the solution of the single-impurity Anderson model (SIAM) [45, 46]. The Slater-Condon integrals corresponding to the L-edge X-ray spectroscopy are obtained with the corresponding modified basis sets, where the core-hole is placed at the respective levels. After solving the SIAM, these integrals and the hybridization functions are used to obtain the X-ray spectra. The double-counting corrections used in this calculation are formulated using the concept of charge-transfer energy [45].

For this calculation, the local Green's function is constructed with the projection of a single particle Green's function from the lattice momentum to the impurity site within a particular energy window. The inversion of the local Green's function relating the hopping parameters from impurity to bath is used to calculate the hybridizations of the projected Fe-$3d$ orbitals with the ligands. After discretization of the continuous hybridization function, the resulting bath states are used in the exact diagonalization scheme, as implemented in the impurity model [47]. From a combination of the DFT-derived single particle Hamiltonian, the Coulomb interaction terms between the impurity $3d$ orbital and relevant bath states and a $2p$ core spin-orbital state, we have constructed the SIAM. The strong spin-orbit coupling of the core states leads to a splitting of the $L_{2,3}$ edges of the absorption spectrum.

**D. Spin-dynamical calculation: procedural details**

With the first-principles extracted intersite exchange parameters, we have dynamically solved the Landau-Lifshitz-Gilbert (LLG) equation to extract the dynamical aspects of the magnetic ground states. More details of this calculation are described in the respective sections. In our simulation, we bring the system to its ground state by performing one million Monte Carlo steps by using Heat bath algorithm. During this process, we continuously monitor the energy of the system to ensure that it stabilizes in its ground state. The convergence criterion is defined such that the changes in the system's energy must be smaller than a threshold of $10^{-3}$ mRy. Once this condition is satisfied, we consider the energy fully converged and the system is equilibrated in its ground state.

After achieving this equilibrium, we proceed with another one million time-step classical spin dynamics simulation, which is governed by the Landau-Lifshitz-Gilbert (LLG) equation. This simulation allows us to investigate the dynamical behavior of the spin system. Specifically, using this spin dynamics simulation, we have obtained the dynamical structural factors of the system, which provide insight into the spatial distribution of the spin-spin correlations. Additionally, we also extract the adiabatic magnon spectrum by using Linear Spin Wave Theory, which characterizes the collective spin excitations and provides valuable information about the magnetic properties of the system.

## III. STRUCTURAL DETAILS AND DYNAMICAL CORRELATIONS

One single layer of the quasi-2D F4GT system constitutes an intricate network of the two inequivalent sites of Fe, *viz*. $Fe_1$ and $Fe_2$, with Ge, which is encapsulated within the outermost Te layers. The consecutive $Fe_1$-$Fe_2$ dimers are distributed across the Ge-atomic plane following an inversion symmetry. F4GT was designed to avoid a high concentration of Fe-atoms on the Ge-plane and thus the middle Ge plane is wrapped within the closely spaced top and bottom $Fe_2$ layers [19,33]. This trilayer formed by $Fe_2$ and Ge is encapsulated within the top and bottom layers formed by $Fe_1$. Thus, the z-directional stacking of one monolayer of F4GT is composed of the layered assembly Te-$Fe_1$-$Fe_2$-Ge-$Fe_2$-$Fe_1$-Te, having two different layers for each of $Fe_1$ and $Fe_2$. Assembly of such quasi-2D monolayers, stacked along the z-direction by mirror reflections, constitutes the bulk structure. Fig. 1(a) depicts the layered network of $Fe_1$ (red) and $Fe_2$ (blue) revealing their dimerized nature and the respective z-directional stacking, which belongs to the space group *R3-m* (no. 166).

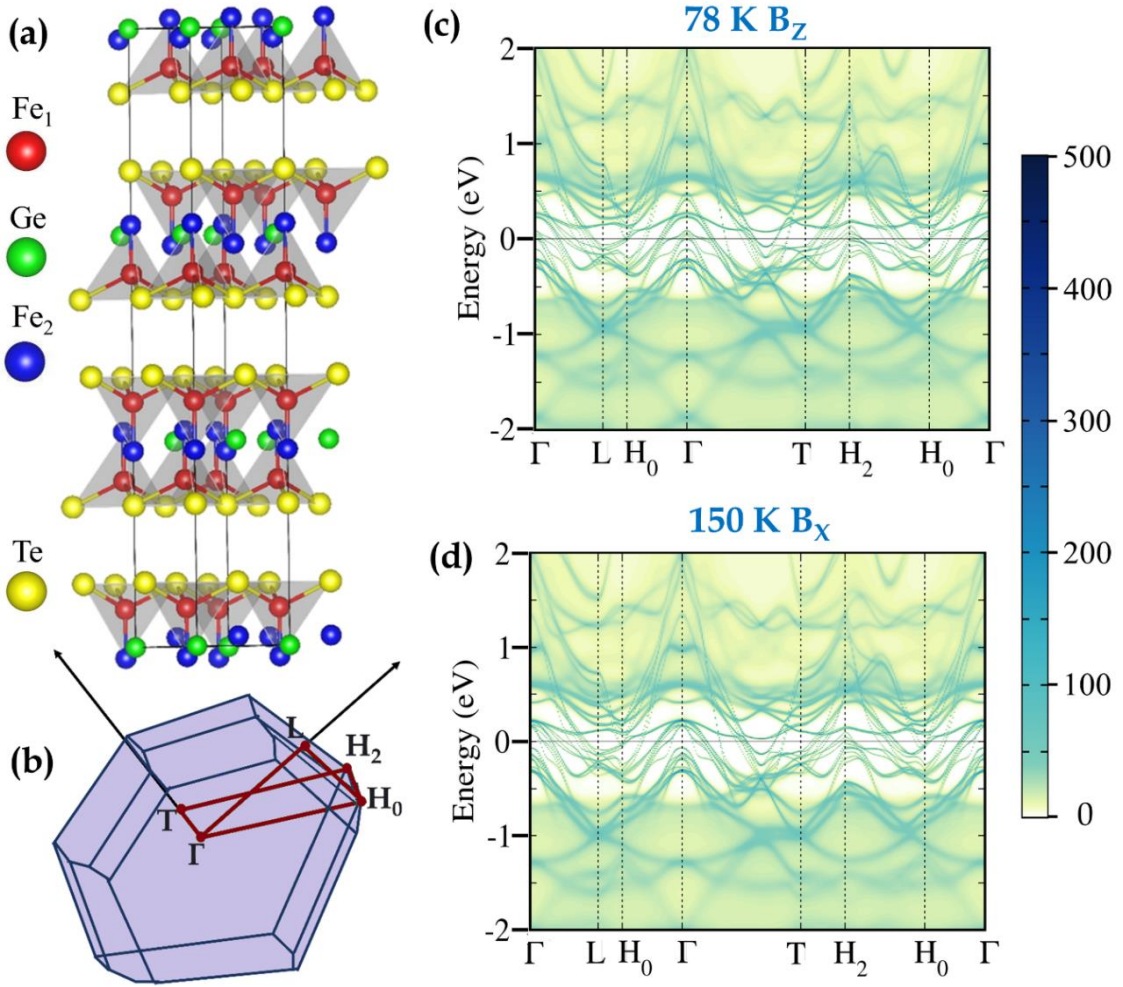

**FIG. 1:** (a) The crystal structure of F4GT with both symmetry inequivalent sites of Fe. (b) The Brillouin zone with the high-symmetry paths used in the spectral intensity plots. The electronic spectral function for $Fe_4GeTe_2$ computed using the spin-polarized fully relativistic DFT+DMFT+SOC method at (c) T = 78 K, magnetization along z-axis (condition 1) and (d) T = 150 K, magnetization along x-axis (condition 2).

In F3GT, specific heat measurements reveal a significant value of the Sommerfeld coefficient,

indicating an enhancement of the quasiparticle mass [48]. For F4GT, the complex nature of the magnetothermal and magnetoresistive results [30, 31] prompts the need for an accurate representation of the local electronic environment [48,49]. Such requirements were also contemplated for the monolayers of the FnGT systems [50].

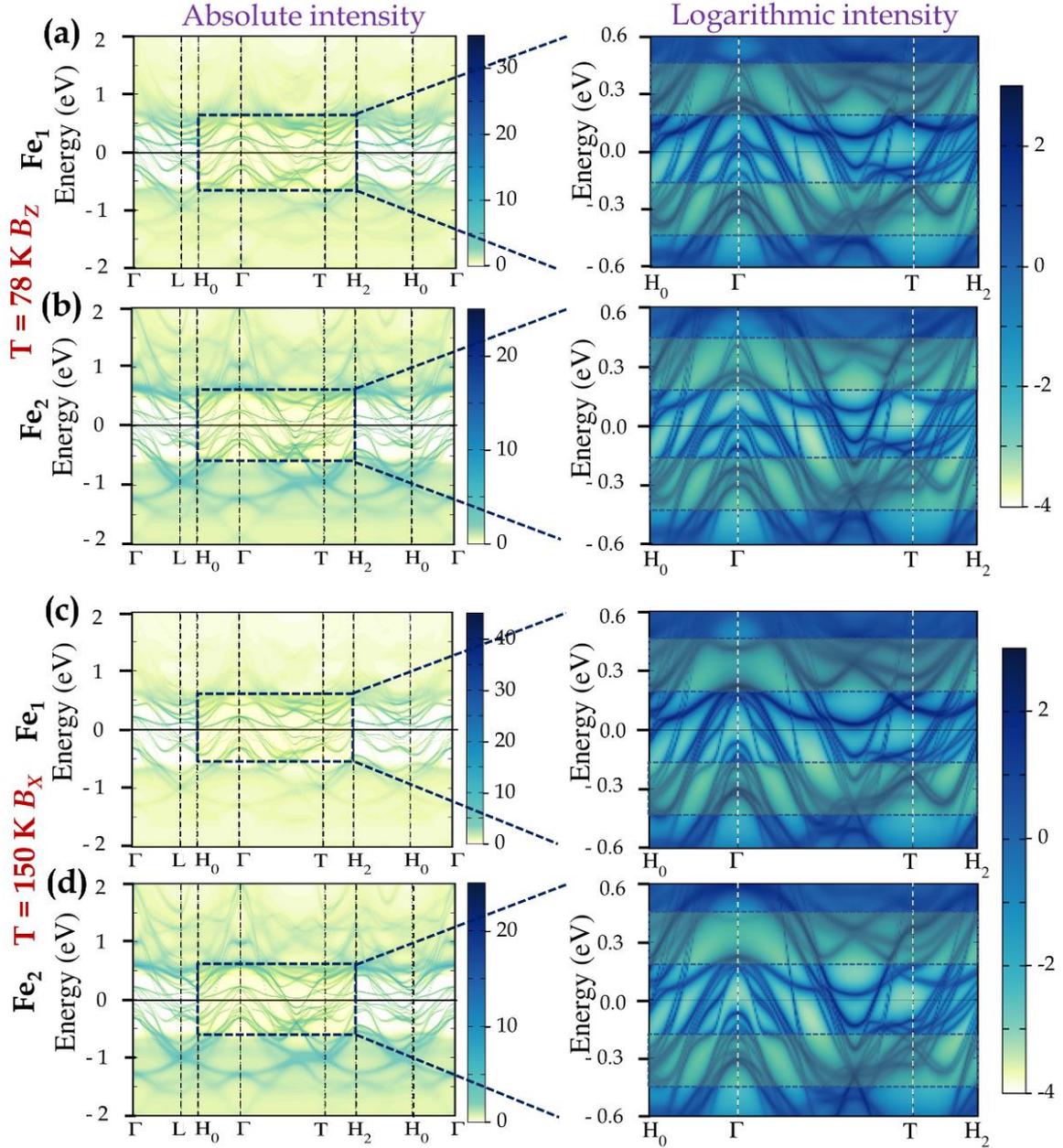

FIG. 2. The electronic spectral function computed using DFT+DMFT+SOC and projected onto the 3$d$ states of Fe$_1$ and Fe$_2$ at (c-d) T = 78 K with magnetization direction along $z$-axis (condition1) and (e-f) T = 150 K with magnetization direction along $x$-axis (condition 2). The next column presents the spectral intensities in a logarithmic scale for a shorter energy and $k$-window. The highlighted portions describe the band-crossing and band-degeneracy (see text).

In our study, we have treated the dynamical electronic correlations of F4GT by using a combined DFT+DMFT approach incorporating SOC, where the self-consistent solution of the effective Anderson impurity model is obtained via the fully relativistic spin-polarized T-matrix fluctuation exchange (SPTF) solver [37, 51-52], as described in section III. Prior magnetic measurements have indicated that for $T > T_{SRT}$ (~ 100 K), the MA evolves from an easy out-of-plane axis to an easy-plane behaviour [29]. We have modelled this experimental scenario in our DFT+DMFT+SOC calculations with the help of two different thermomagnetic conditions, *viz.* (1) at 78 K ($< T_{SRT}$) with the direction of magnetization along the local *z*-axis and (2) at 150 K ($> T_{SRT}$) with the direction of magnetization along the local *x*-axis. Throughout this work, we will refer to these two thermomagnetic conditions as conditions 1 and 2 and the same are referred to in the figures as 78 $K\ B_z$ and 150 $K\ B_x$ respectively. Fig. 1 (c) – (d) depict the total electronic spectral functions for both of these conditions.

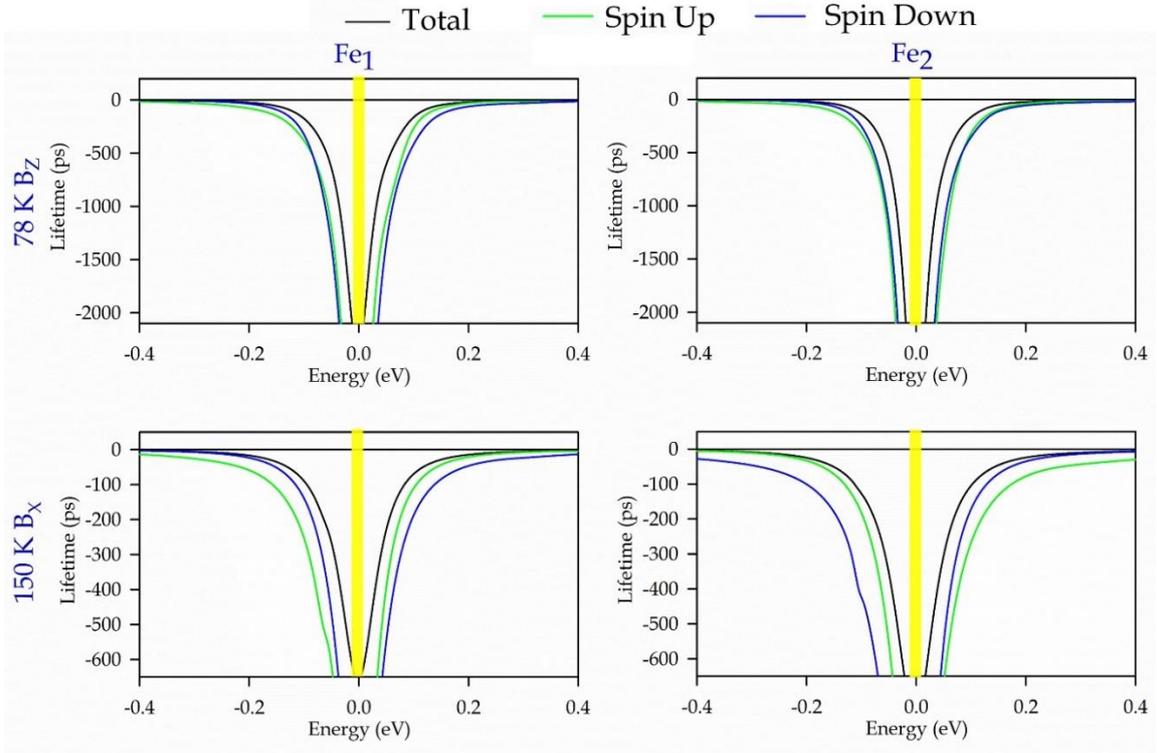

**FIG. 3:** Calculated lifetimes of the quasiparticles from $Fe_1$ and $Fe_2$ of $Fe_4GeTe_2$ computed using the DFT+DMFT+SOC method at T = 78 K, magnetization along *z*-axis (condition1) and at T = 150 K (condition 2), magnetization along *x*-axis.

In Fig 2, the left column of 2(a)-(d) represents the $3d$-projected electronic spectral functions for $Fe_1$ and $Fe_2$, plotted for both conditions described above along the high-symmetry path, as shown with red lines in the Brillouin zone shown in Fig. 1(b). The distinct line-like features of the $k$-resolved spectral functions around $E_F$ and their broadenings away from $E_F$, as seen in both Figs 1 and 2, originate from the finite lifetimes of the respective quasiparticles for both types of Fe.

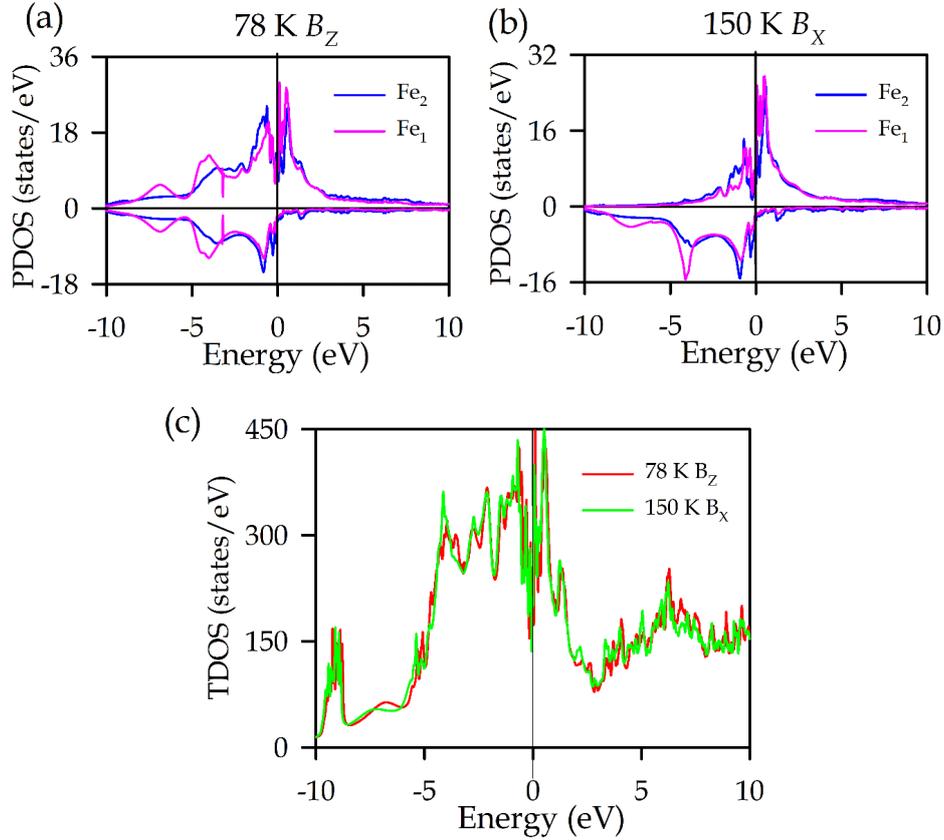

**FIG. 4:** Partial density of states for $Fe_1$ and $Fe_2$-$3d$ states of $Fe_4GeTe_2$ computed using the DFT+DMFT+SOC method at (a) T = 78 K, magnetization along $z$-axis (condition 1) and at (b) T = 150 K, magnetization along $x$-axis (condition 2). (c) Total density of states of $Fe_4GeTe_2$ computed using the DFT+DMFT+SOC method at T = 78 K, magnetization along $z$-axis (condition 1) and at T = 150 K, magnetization along $x$-axis (condition 2).

A quatitative representation of the effect of the strong electron-electron correlation at both conditions for the $3d$-electrons of both types of Fe can be obtained from the orbital-projected and total effective masses of the carriers. These effective masses, as extracted from the DFT+DMFT+SOC calculations, are presented in Table I. As can be seen from the table, the effective masses of the $Fe_1$ and $Fe_2$-3$d$ electrons are around 1.5-1.6, which indicates their moderately correlated nature.

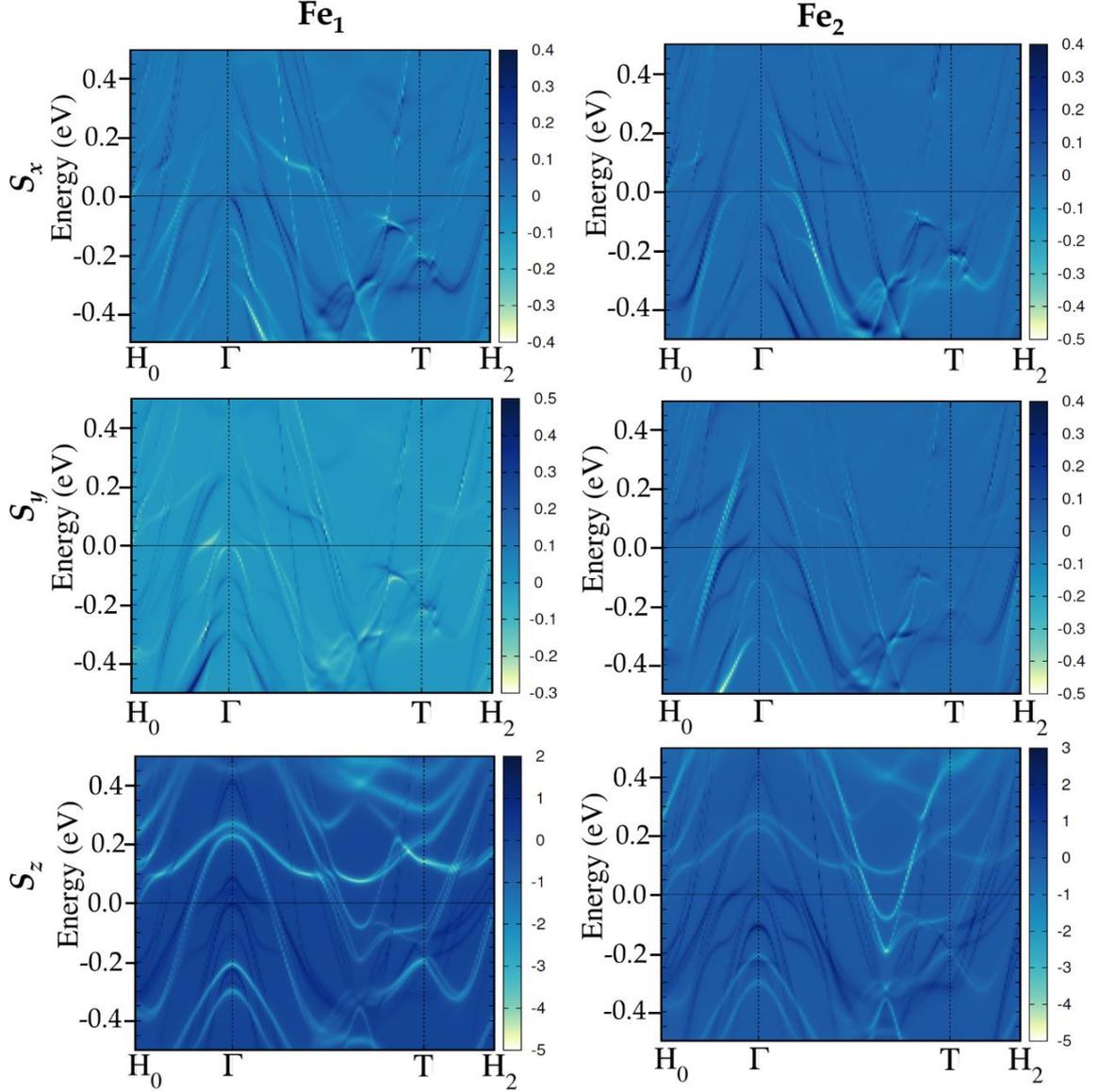

**FIG. 5:** Spin-component-projected partial spectral functions for Fe-3$d$ states of $Fe_4GeTe_2$ computed using the DFT+DMFT+SOC method at T = 78 K, magnetization along $z$-axis (condition1) for limited region of the BZ within a zoomed-in energy window. The associated colour scale is a measure of the quasiparticle lifetimes.

Comparison of the type-projected values of the total effective masses also reveal that the $Fe_1$-3$d$ electrons have marginally higher masses than those of the $Fe_2$-3$d$ ones. In Fig. 3, the lifetimes (τ) of Fe$_1$ and Fe$_2$, as calculated from the inverse of the imaginary part of the self-energy as per equation (9) of section III, are plotted. According to the DFT + DMFT formulation, the imaginary part of self-energy tends to zero as the energy approaches $E_F$. Therefore, in Fig. 3, both of the lifetimes blow up to very large values within a band-offset of 0.001 eV, as approximately shown by the yellow highlighted region. Away from $E_F$, the gradual decay of the lifetimes is related to the broadening of the spectral functions.

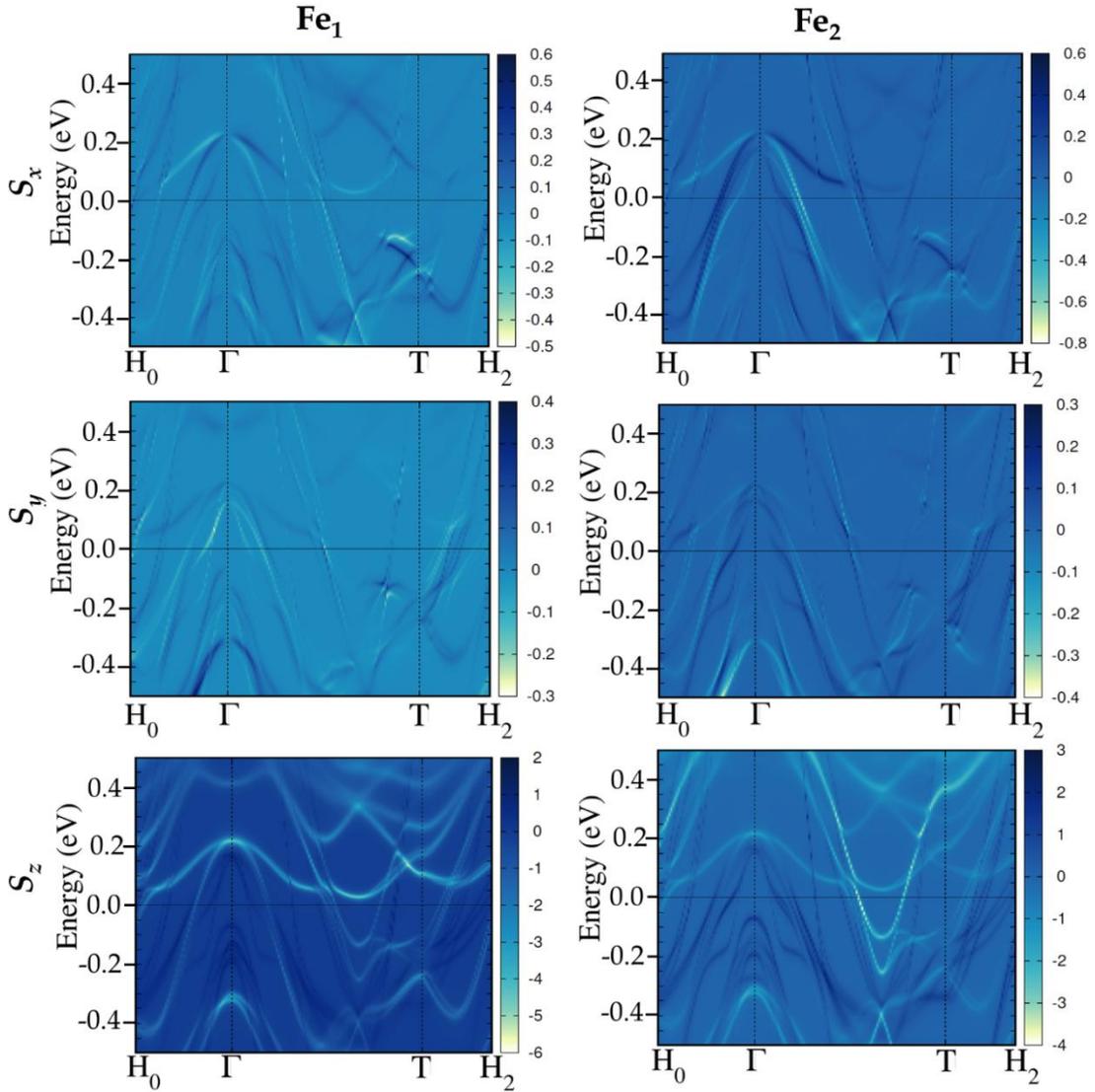

**FIG. 6:** Spin-component-projected partial spectral functions for Fe-3$d$ states of Fe$_4$GeTe$_2$ computed using the DFT+DMFT+SOC method at T = 150 K, magnetization along $x$-axis (condition 2) for limited region of the BZ within a zoomed-in energy window. The associated colour scale is a measure of the quasiparticle lifetimes.

Interestingly, the low-energy spectral functions reveal a strong influence from the electronic temperature and the symmetry breaking associated with the magnetization direction. In Figs. 2(a)-(d), the right column presents the corresponding spectral functions in a logarithmic intensity scale, plotted in a smaller energy window, beyond which there is significant lifetime broadening of the quasiparticle states. This energy window is marked by a black-dashed rectangle around $E_F$ along the high-symmetry path $H_0$-$\Gamma$-T-$H_2$ in the left column of Fig. 2(a)-(d). In the logarithmic plots, two different energy-ranges are highlighted, indicating that these are the spectral regions with the most prominent electronic footprints, which originates from the differences in conditions 1 and 2. A close comparison of these absolute and logarithmic intensity plots in these three regions reveals the following details:

(A) The first region is around $E_F$ and the others are around ± 0.3 eV, as seen in the highlighted strips. For the first one, while comparing the near-$E_F$ bands at the $\Gamma$-point for both conditions, we observe a splitting of those bands for condition 2, opening a pseudo-gap of ~ 0.2 eV near $\Gamma$-point. The opening of this pseudogap beyond $T_{SRT}$ can be attributed to the lowering of the symmetry for condition 2 due to the transition of the magnetization direction from the out-of-plane to the in-plane geometry. The symmetry operations available for condition 1 are the identity operation, one three-fold clockwise rotation around [001], one three-fold anticlockwise rotation around [001], inversion around [001], one three-fold anticlockwise rotation around [001] + inversion around [001], another three-fold clockwise rotation around [001] + inversion around [001]. Thus, altogether there are six symmetry operations. On the other hand, the symmetry operations available for condition 2 are the identity operation, one two-fold anticlockwise rotation around [100], one inversion around [001] and another three-fold anticlockwise rotation around [100] + inversion around [001]. Therefore, in condition 2, only four symmetry operations are available. Due to this lowered symmetry, beyond $T_{SRT}$, opening of this pseudo-gap-like feature will also be evident from the near-$E_F$ lowering of the partial and total density of states (DOS) as plotted in Fig. 4(a)-(c). Thus, the occurrence of this pseudogap can be considered as the electronic footprint of SRT. This footprint is also verified to be present within simple DFT and therefore is not related to the DMFT-induced band renormalization.

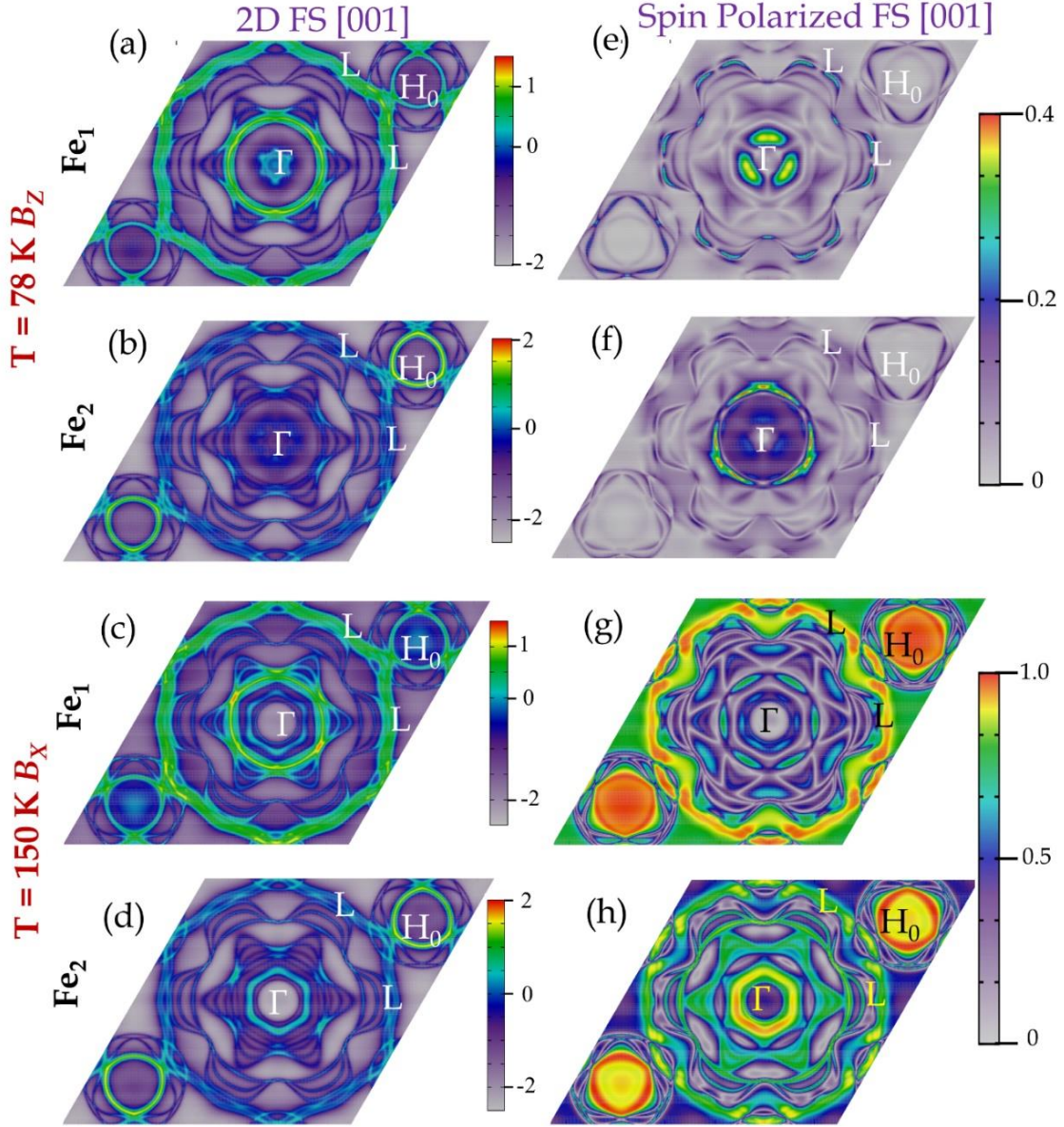

FIG. 7. (a-d) 2D Fermi surfaces for the [001] plane in the logarithmic intensity scale for 3$d$ states of Fe$_1$ and Fe$_2$ at (a-b) T = 78 K with magnetization direction along $z$-axis (condition1) and (c-d) T = 150 K with magnetization direction along $x$-axis (condition 2). Spin-polarized FS in the absolute intensity scale for the Fe$_4$GeTe$_2$ computed using the DFT+DMFT+SOC method at (e,f) T = 78 K, magnetization along local $z$-axis and (g,h) T = 150 K, magnetization along $x$-axis. The adjacent color scale represent the degree of spin-polarizations.

(B) This splitting of bands at the Γ-point from condition 1 to 2 has additional two-sided impacts in the second energy-window around ± 0.3 eV along the high symmetry path Γ-T. First, at +0.3 eV, near Γ-point, there is a shift of bands towards lower energy, leading to several almost degenerate bands at ~ 0.2 eV. Second, for condition 2 and along Γ-T, one set of linearly dispersive bands cross each other at ~ 0.3 eV.

(C) Similarly, around -0.3 eV, the split bands at the Γ-point in condition 1 shift downwards and become degenerate for condition 2. Moreover, along Γ-T, another pair of linearly dispersive bands cross each other at ~ -0.3 eV.

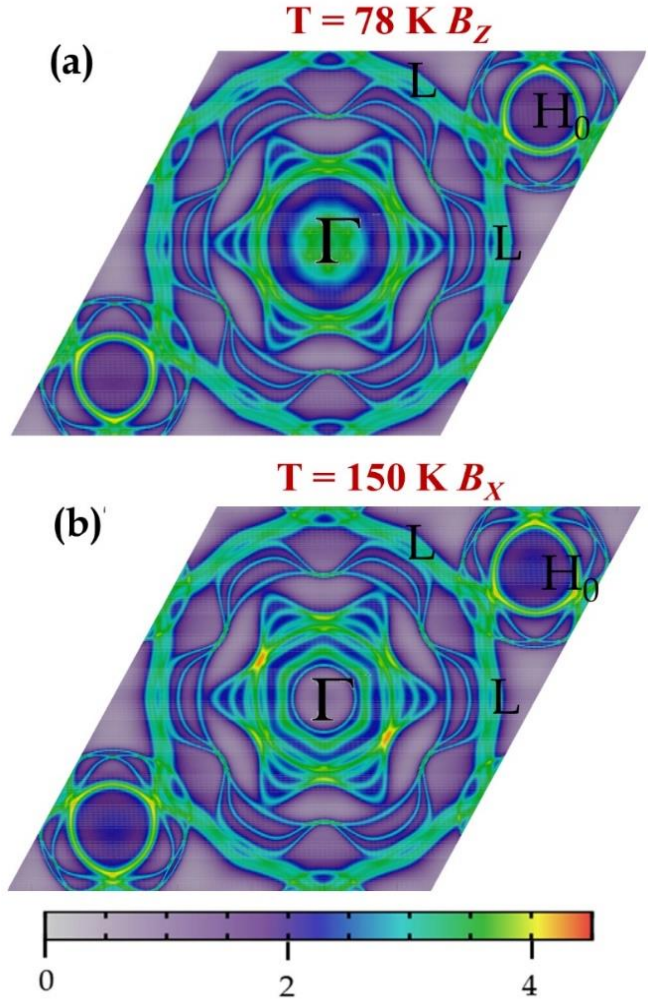

**FIG. 8:** (a,b) The resultant 2D FS Fe$_4$GeTe$_2$ computed using the spin-polarized fully relativistic DFT+DMFT+SOC method at (a) T = 78 K, magnetization along *z*-axis (condition 1) and (b) T = 150 K, magnetization along *x*-axis (condition 2).

These linearly dispersive band-crossings away from $E_F$, albeit having lesser importance for stoichiometric bulk systems, can impart an important contribution for the non-stoichiometric electron or hole-doped systems with shifted $E_F$. In its ground state with SOC-induced non-collinear magnetic configuration, these impacts of band splittings and crossings are seen to have more contributions from the $z$- and $x$-projections of spins for conditions 1 and 2 respectively for both $Fe_1$ and $Fe_2$, as seen in Figs. 5 and 6. As was also pointed out earlier in this article that prior experiments of anomalous Hall effects have indicated the presence of topological band crossings near $E_F$ [29-31]. Keeping the non-stoichiometry of the experimental systems in mind, the presence of the band-crossings pointed out in our study, may be useful for future experimental studies.

**Table I:** The effective mass ($m^*/m_{DFT}$) from DFT+DMFT (quasiparticle weight as calculated from the imaginary part of self-energy).

| Type | T (K) | Spin | $d_{z^2}$ | $d_{x^2-y^2}$ | $d_{xz}$ | $d_{yz}$ | $d_{xy}$ | Total |
|---|---|---|---|---|---|---|---|---|
| Fe1 | 78 K | Spin down | 1.481 | 1.481 | 1.461 | 1.461 | 1.481 | 1.620 |
| | | Spin up | 1.769 | 1.792 | 1.831 | 1.831 | 1.792 | |
| | 150 K | Spin down | 1.503 | 1.490 | 1.506 | 1.481 | 1.490 | 1.623 |
| | | Spin up | 1.760 | 1.757 | 1.766 | 1.814 | 1.795 | |
| Fe2 | 78 K | Spin down | 1.332 | 1.412 | 1.440 | 1.440 | 1.412 | 1.506 |
| | | Spin up | 1.572 | 1.595 | 1.680 | 1.680 | 1.595 | |
| | 150 K | Spin down | 1.381 | 1.342 | 1.404 | 1.426 | 1.381 | 1.485 |
| | | Spin up | 1.545 | 1.550 | 1.550 | 1.650 | 1.655 | |

Figs. 7(a)-(d) present the 2D FS projected across the [001] plane in the logarithmic intensity scale for $3d$ states of $Fe_1$ and $Fe_2$ under both conditions 1 and 2 respectively. Here, higher

intensity of $Fe_1$-$3d$ projected sections in comparison to the $Fe_2$-$3d$ projections has its origin in the higher effective masses of $Fe_1$-$3d$ electrons near $E_F$. A closer inspection near the $\Gamma$-point elucidates the effect of band splitting as encountered in condition 2 by displaying a clear $\Gamma$-centred hole pocket, which is otherwise blurred for condition 1. The $H_0$-centred hole pockets for $Fe_1$ in the first condition is smeared for the second one, due to the splitting and consequential shifting of bands as a result of lowered symmetry for condition 2. All of these effects are also obvious in the resultant FS as plotted in Fig. 8.

**Table II:** Spin moment and orbital moment for Fe in $Fe_4GeTe_2$ computed using the DFT+U+SOC and DFT+DMFT+SOC methods.

| | | Calculated moment ($\mu_B$) | | | | Experimental values of the moments | |
|---|---|---|---|---|---|---|---|
| Method | $Fe_1$/$Fe_2$ | T = 78 K $B_z$ | | T = 150 K $B_x$ | | | |
| | | Spin | Orbital | Spin | Orbital | Spin | Orbital |
| DFT+U+SOC | $Fe_1$ | 2.92 | 0.14 | 2.90 | 0.22 | 1.7-2.2 $\mu_B$/Fe | 0.05 $\mu_B$/Fe |
| | $Fe_2$ | 2.57 | 0.18 | 2.56 | 0.15 | Ref. [26, 29, 30, 33] | Ref. [33] |
| DFT+DMFT+SOC | $Fe_1$ | 2.66 | 0.07 | 2.64 | 0.14 | | |
| | $Fe_2$ | 2.20 | 0.05 | 2.19 | 0.09 | | |

Figs. 7(e)-(h) depict the spin-polarized FS for the [001] surface of F4GT, where the projected quantity is the in-plane linear superposition of the $x$ and $y$ components of the spin-magnetic moment. The adjacent color scale signifies the degree of spin polarization. For condition 1, there are negligible in-plane polarizations, supporting the fact that the magnetization vector for $T < T_{SRT}$ is along the out-of-plane ($z$) axis. On the contrary, for condition 2 with $T > T_{SRT}$, the [001] plane is highly spin-polarized for both $Fe_1$ and $Fe_2$, indicating the highest degree of polarization near $H_0$. This result also corroborates with experiment, where the in-plane magnetization is much higher than the out-of-plane one. Fig. 9 presents the spin-component projected FS in the [001] plane, where projections along the chosen magnetization directions

are seen to possess more intensities.

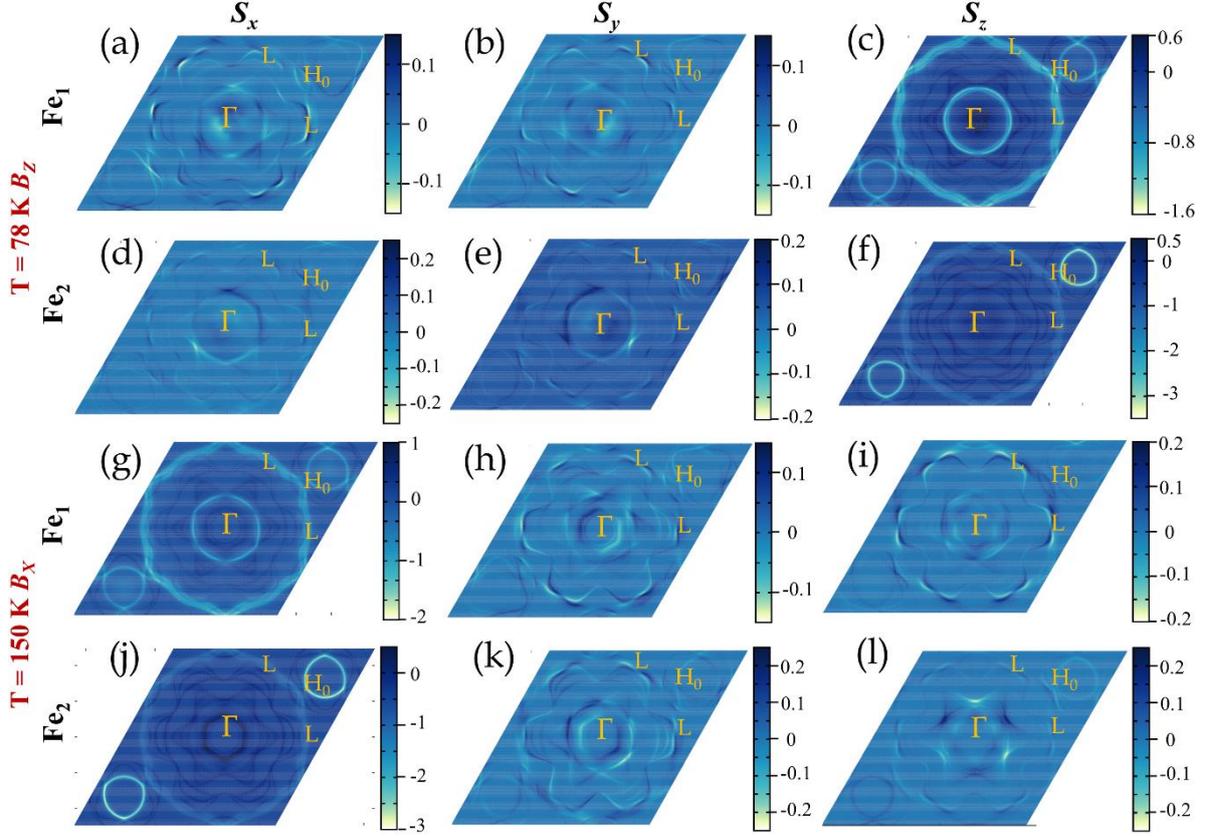

**FIG. 9:** Spin-component-projected 2D FS for $Fe_1$ and $Fe_2$ of $Fe_4GeTe_2$ computed using the spin-polarized and relativistic DFT+DMFT+SOC method at (a-f) T = 78 K, magnetization along local *z*-axis (condition 1) and (g-l) T = 150 K, magnetization along *x*-axis (condition 2). The colour bar in side panel represents the logarithmic intensity.

## IV. X-RAY SPECTRAL PROPERTIES: DFT + DMFT+ MLFT

Prior experimental studies of the x-ray spectral properties like x-ray absorption spectroscopy (XAS) and x-ray magnetic circular dichroism (XMCD) of the two Fe sites of F4GT have indicated an isotropic orbital occupation and a lower MA than its other FnGT counterparts [33]. However, it is difficult to experimentally segregate the contributions from the two symmetric sites of Fe towards its x-ray spectral properties. In Figs. 10(a)-(b), we present the calculated XAS and XMCD spectra of $Fe_1$ and $Fe_2$ after coupling the DFT+DMFT+SOC with the multiplet ligand field theory (MLFT) as described in the Ref. [45,47]. In this calculation, we considered the two collinear Fe moments as ferromagnetically coupled, since an AFM coupling results in an almost vanishing XMCD signal which is inconsistent with the experiments. The $L_3$ and $L_2$ edges represent transitions from the $2p_{1/2}$ and $2p_{3/2}$ to the $3d$ levels of Fe, respectively. As can be seen from the green dotted lines in Fig.10(a) and (c), the computed $L_3$ and $L_2$ XAS spectra match closely with the experimental data [33], both concerning position of the peaks as well as the relative intensity (the branching ratio).

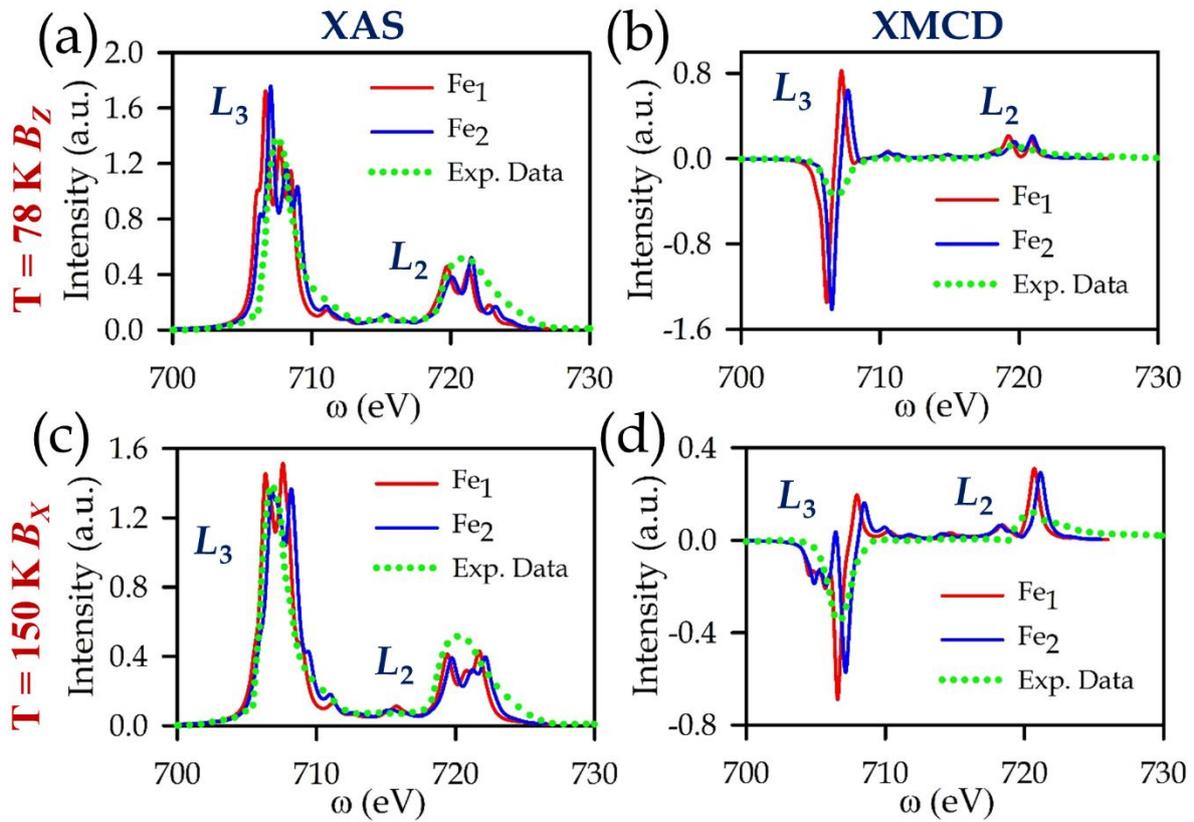

FIG. 10. XAS and XMCD $L_{2,3}$-edge spectra for $Fe_1$ (red) and $Fe_2$ (blue) for (a-b) condition 1 and (c-d) condition 2 respectively (see text). The green dotted lines present the experimental data of reference [33].

The calculated theoretical spectra show a clear distinction between $Fe_1$ and $Fe_2$ with the prominent peaks that reflect the multi-configurational aspect of the excitation (although not shown, a simple plot of the unoccupied DOS does not reflect the experimental spectrum). Moreover, while comparing the projected FS corresponding to these two different magnetization directions, $Fe_2$ (Fig. 7(f)) and $Fe_1$ (Fig. 7(g)) are respectively contributing more towards spin-polarization in the [001] plane. This corroborates with the fact that the experimental XMCD peak positions match more closely with the spectra corresponding to $Fe_2$ and $Fe_1$ for conditions 1 and 2 respectively [33].

In Table II, we compare the moments calculated by using DFT + (static) Hubbard U + SOC and DFT+DMFT+SOC calculations, which reveal the closeness of the experimentally extracted magnetic moments for the later one, as presented in last column of the same Table. These spin magnetic moments for $Fe_1$ and $Fe_2$ from DFT+DMFT+SOC are ~ 2.6 and 2.2 $\mu_B$ respectively, which are close to the experimental values ranging from 1.7- 2.2 $\mu_B$ [26, 29, 30, 33]. The corresponding orbital moment values, as seen in the same Table, resemble the experimental orbital moments of ~ 0.05 $\mu_B$, as extracted from the XMCD measurements [33].

## V. INTERSITE EXCHANGE INTERACTIONS: DFT + DMFT + LKAG

In F4GT, the intricate details of the magnetic interactions and their interplay with the MA remain an open question. The calculated values of MA with DFT+DMFT+SOC reveal that the low-temperature higher value (0.13 mRy/Fe) of the easy-axis anisotropy gradually decreases with increasing temperature and beyond $T > T_{SRT}$, the MAE remains almost constant with a lower value of ~ 0.047 mRy/Fe. By combining the DFT+DMFT+SOC calculations with the LKAG formalism, the intersite exchange parameters of F4GT were calculated, with an aim to describe the low-energy magnetic excitations from an effective spin Hamiltonian [41,43], including terms like symmetric isotropic and anisotropic Heisenberg exchange ($J_{ij}$ and $\Gamma_{ij}$) and the antisymmetric and anisotropic Dzyaloshinskii Moriya ($D_{ij}$) exchange. In this method, the concerned electronic system is mapped onto a generalized classical Heisenberg model as per the equation $H = - \sum_{i \neq j} e_i^\alpha \hat{J}_{ij}^{\alpha\beta} e_j^\beta$, $\alpha, \beta = x, y, z$. Here $e_i$ and $e_j$ are the local spin vectors at the $i$-th and $j$-th site and $\hat{J}$ is the exchange tensor. Different components of the intersite exchange parameters like $J_{ij}$, $D_{ij}$ and $\Gamma_{ij}$ are calculated from $\hat{J}$ by following the equations (11)-(12) as described in the section III. Fig. 11 illustrates the components of these exchange parameters calculated along the magnetization directions in the global coordinate system.

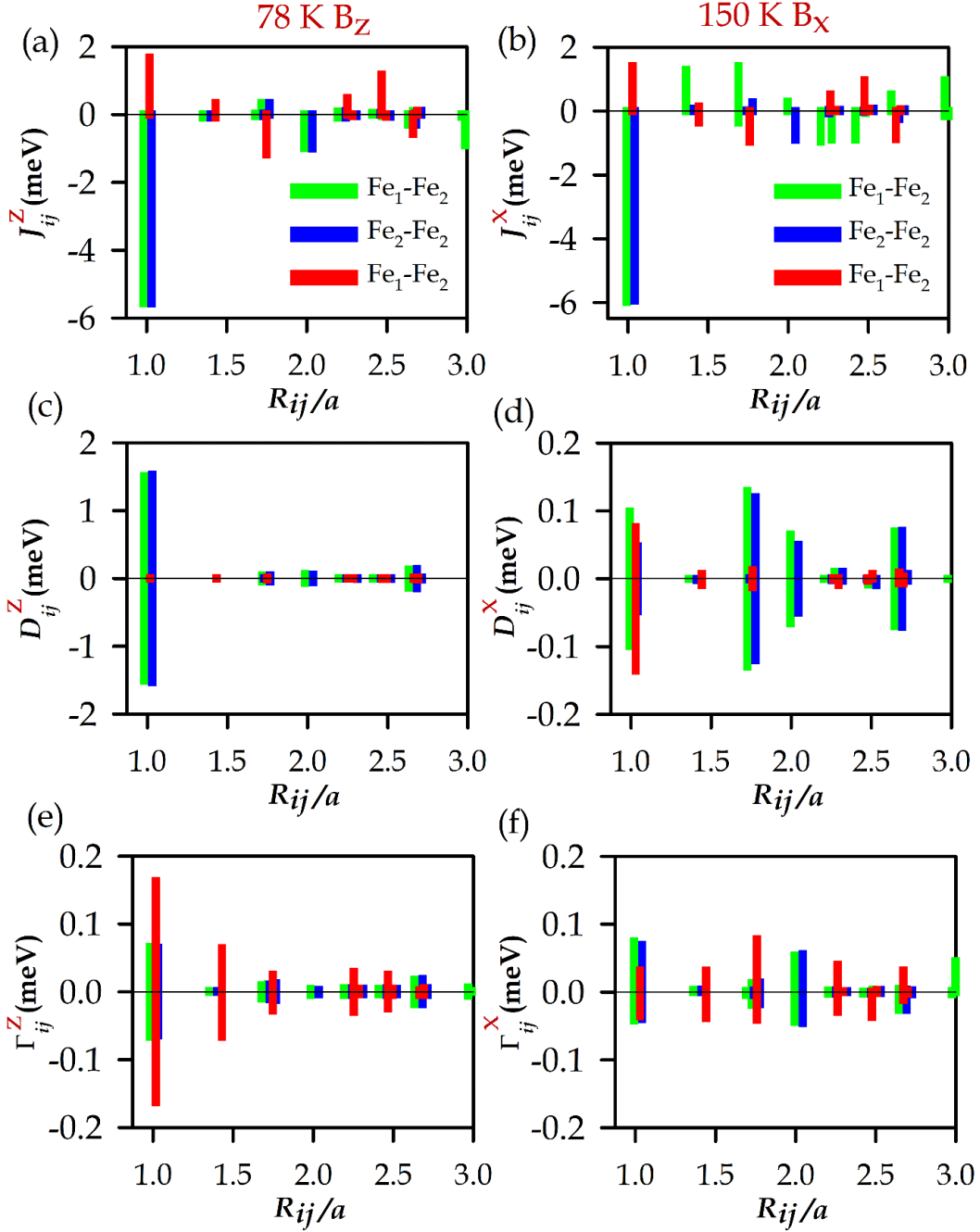

FIG.11. (a, c, e) Components of the Heisenberg exchange interaction $J_{ij}^z$, DM interaction $D_{ij}^z$ and $\Gamma_{ij}^z$ of Fe$_4$GeTe$_2$ calculated with DFT+DMFT+SOC for condition 1; (b, d, f) components of the Heisenberg exchange interaction $J_{ij}^x$, DM interaction $D_{ij}^x$ and $\Gamma_{ij}^x$ of Fe$_4$GeTe$_2$ calculated with DFT+DMFT+SOC for condition 2 corresponding to the different pairwise interactions between Fe$_1$ and Fe$_2$. These intersite parameters are plotted with respect to the NN distance (scaled by lattice parameter) between the Fe-atoms.

Thus, for condition 1 and 2, Fig 11(a) and (b) depict the $J_{ij}^{z}$ and $J_{ij}^{x}$ parameters corresponding to the three possible pairwise Fe-Fe interactions, *viz*. $Fe_1$-$Fe_1$, $Fe_2$-$Fe_2$ and $Fe_1$-$Fe_2$ interactions for both conditions 1 and 2 with respect to the nearest-neighbour (NN) distances scaled by the lattice parameters [53,54]. The nature of these interactions reveals the following intriguing aspects:

(1) the first NN exchanges for both $Fe_1$-$Fe_1$ and $Fe_2$-$Fe_2$ interactions are negative for both thermomagnetic conditions 1 and 2 indicating the strong antiferromagnetic (AFM) nature of the first NN interactions between Fe-atoms of similar symmetric types;

(2) the first NN interactions for the $Fe_1$-$Fe_2$ pairs, on the contrary, is weakly ferromagnetic (FM) for both conditions;

(3) the nature of variation of the $J_{ij}$ components for all these three types of pairwise interactions with atomic distances follows an almost oscillatory Rudderman-Kittle-Kasuya-Yoshida (RKKY)-like pattern. Such a pattern indicates the impact of the interaction of the local magnetic moments of $Fe_1$ and $Fe_2$ with the itinerant electronic background of the system [55]. The RKKY-like oscillatory behaviour of the $J_{ij}$ parameters becomes more evident from Fig. 12, where we plot $J_{ij} \times R^3$ as a function of distance (R) for three types of pairwise interactions. Although the scaling factor should be slightly reduced to account for the quasi-2D nature of F4GT [56], our analysis is sufficient to identify the signature of the RKKY behavior;

(4) The overall $J_{ij}$ values for $Fe_1$ and $Fe_2$ are similar for both conditions. However, the *z*-component of the DM parameter for condition 1 is one order of magnitude larger than the *x*-component of the same for condition 2, as plotted in Fig. 11(c) and (d). This happens due to the temperature-induced weathervane effect leading to a disruption of the magnetic chirality at higher temperature [57]. Overall, the values of the DM-parameters are quite high leading to a high $D/J$ ratio at lower temperature;

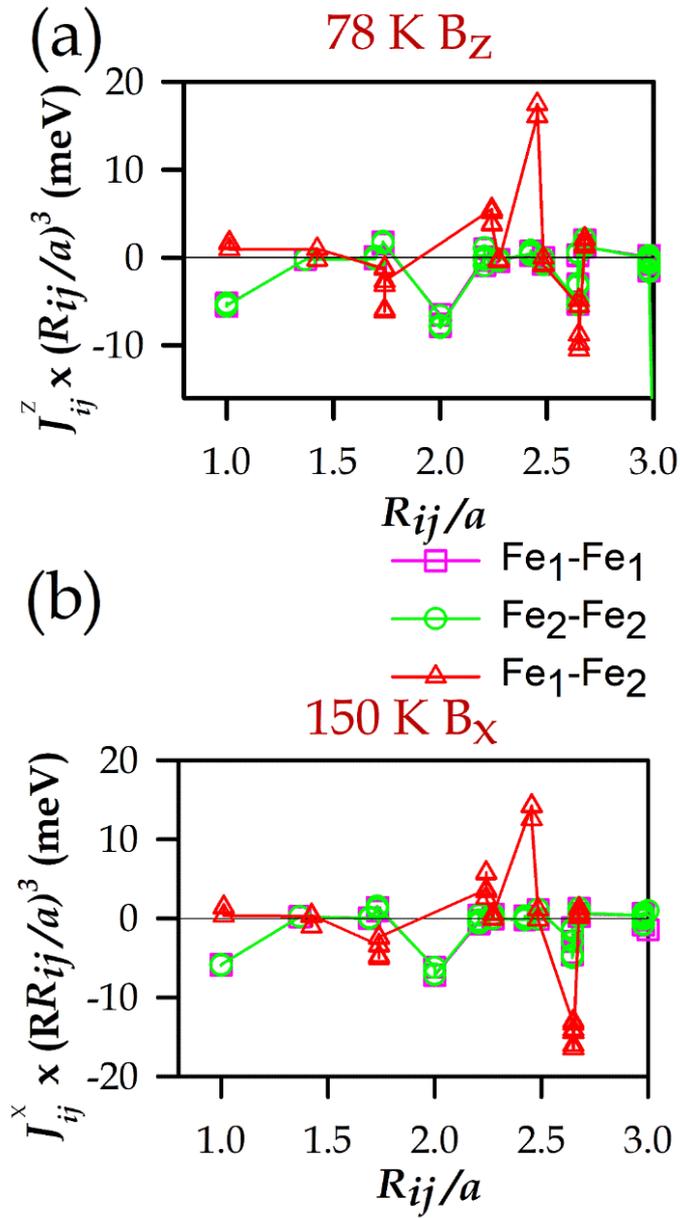

**FIG. 12:** The Heisenberg exchange interaction $J_{ij} \times R^3$ ($R$ = NN distance) for pairwise interactions of $Fe_1$ and $Fe_2$ of $Fe_4GeTe_2$ calculated using a combination of DFT+DMFT+SOC and LKAG at T = 78 K, magnetization along $z$-axis (condition 1) and at T = 150 K, magnetization along $x$-axis (condition 2). The oscillatory behaviour implies an RKKY-like interaction.

(5) for both conditions, there are non-zero DM and $\Gamma$ interactions present in the system, with their values having opposite signs for the same Fe-Fe distance, implying the possible presence of a chiral magnetic ground state [57]. These interactions are also highly anisotropic, where atoms at different directions bearing the same NN distance are having different DM and Γ interactions;

(6) whereas the $D_{ij}$ values for condition 1 and for the first NN are almost one order of magnitude higher than the next NN, the same values for condition 2 display a zigzag behaviour.

Experimental studies have not yet provided a complete understanding of the magnetic interactions in F4GT except for the prediction of the presence of complex magnetic interactions within the system [30-32]. Our extracted intersite exchanges and pairwise interactions confirm the suggested complexity, while also emphasizing the differences between the underlying inter- (Fe$_1$-Fe$_2$) and intra-type (Fe$_1$-Fe$_1$ and Fe$_2$-Fe$_2$) magnetic interactions, pointing towards the tendency to obtain a chiral magnetic ground state.

## VI. MAGENTOTHERMAL PHASE DIAGRAM: SPIN DYNAMICAL SIMULATIONS

With the extracted values of the intersite exchange parameters, the generalized bilinear effective Hamiltonian is constructed with terms like: $\mathcal{H}_{mod}^{i,j} = -J_{ij}\boldsymbol{e}_i \cdot \boldsymbol{e}_j - \boldsymbol{D}_{ij}\boldsymbol{e}_i \times \boldsymbol{e}_j - \boldsymbol{e}_i\Gamma_{ij}\boldsymbol{e}_j - \kappa\sum_{k=i,j}(\boldsymbol{e}_k \cdot \boldsymbol{e}_k^r)^2$ [58-61]. Here $\boldsymbol{e}_i$ and $\boldsymbol{e}_j$ are the unit vectors along the direction of the spin moments at the atomic sites $i$ and $j$ and $\boldsymbol{e}_k^r$ is the easy axis, along the arbitrary unit vector $\boldsymbol{r}$. The most general spin-dynamical solution of this Hamiltonian leads to a homogeneous chiral magnetic ground state, where the tensor atomistic moment has the form $\boldsymbol{m}_{n\alpha} = m_\alpha \begin{pmatrix} sin\theta_\alpha cos(\varphi_{n\alpha} + \gamma_\alpha) \\ sin\theta_\alpha sin(\varphi_{n\alpha} + \gamma_\alpha) \\ cos\theta_\alpha \end{pmatrix}$ [62, 63]. Here $\theta_\alpha$ and $\gamma_\alpha$ are the cone and the phase angles, respectively. The azimuthal angle corresponding to the local magnetic moment can be calculated as $\varphi_{n\alpha} = \boldsymbol{q} \cdot \boldsymbol{R}_{n\alpha}$. With the help of atomistic spin dynamics simulations, as implemented in the UppASD software [59, 60], the dynamics of small fluctuations of the spins around the classical direction of the local spin moment are taken into account to obtain the dynamical magnetic ground state. The presence of low-energy collective magnetic excitations or magnons in this system can be analyzed after calculating the Fourier transform of the spin-spin correlation function or the dynamical structure factor as; $S(q,\omega) = \frac{1}{2\pi N}\sum_{i,j} e^{iq(r_i-r_j)} \int_{-\infty}^{\infty} d\tau e^{-i\omega\tau} \langle e_i e_j(\tau) \rangle$, where, $r_i$ is the position vector of the magnetic atoms and $e_i$ is the local spin-vector.

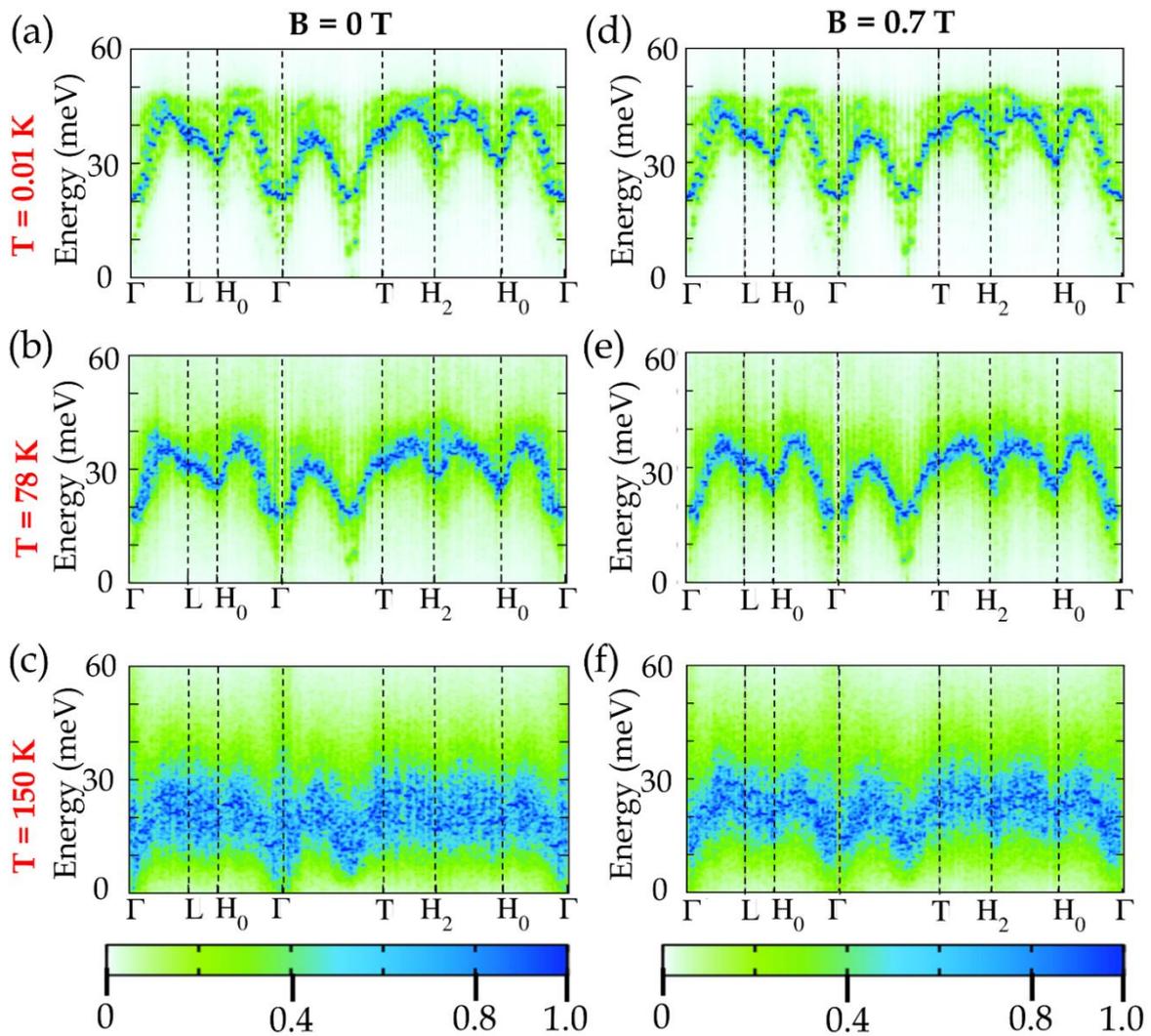

FIG. 13. The resultant dynamical structure factors $S(q,\omega)$ plotted along the high-symmetry paths for F4GT at (a,d) T = 0.01 K, (b,e) T = 78 K and (c,f) T = 150 K with an applied field of (a-c) 0 T and (d-f) 0.7 T.

In Figs. 13(a)-(c), we have plotted the resultant $S(q,\omega)$ values along different high-symmetry directions for three thermal conditions, *viz.* 0.01 K, 78 K and 150 K without any applied field.

In the next column, we have plotted the resultant $S(q,\omega)$ values for the same set of temperatures with an applied field of 0.7 T. The associated colour scale represents the normalized values of $S(q,\omega)$, where the regions with higher intensity (blue in colour) represent the adiabatic magnon dispersion lines.

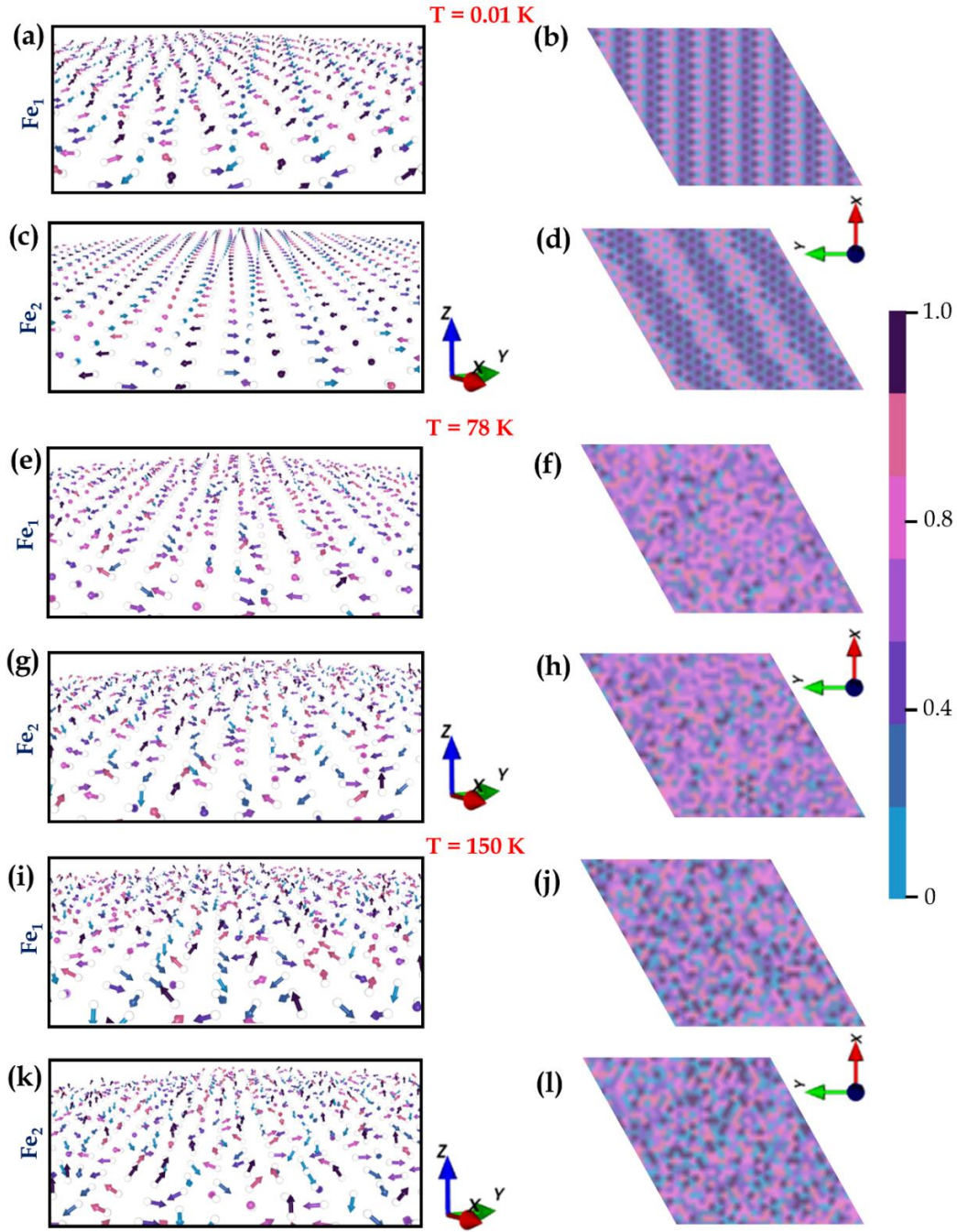

**FIG. 14:** The 3D view of the dynamical ground-state spin-textures (left panel) and the magnetization densities (right panel) at (a-d) T = 0.01 K, (e-h) T = 78 K, and (i-l) T = 150 K, for each atomic type of $Fe_1$ and $Fe_2$. The spin-spirals are aligned along x-direction.

Presence of multiple local minima at Γ and in between Γ-T indicates a complex superposition of chiral orders within the system. The magnetic planar textures and the corresponding magnetization densities corresponding to the Figs. 13(a)-(c) are presented in Fig. 14. Below $T_{SRT}$, the zero-field $S(q,\omega)$ plots do not display much difference with the 0.7 T plots. With increasing temperature, we observed a significant reduction of the magnon energy range. For T = 150 K and B = 0.7 T, the minimas of low-energy spin correlations at Γ and in between Γ-T become more similar (Fig. 13(c)) than their corresponding different values at zero-field (Fig. 13(f)) due to the temperature-induced statistical fluctuations of the spin-orientations and a field-induced opening of a gap at the Γ-point with an applied field.

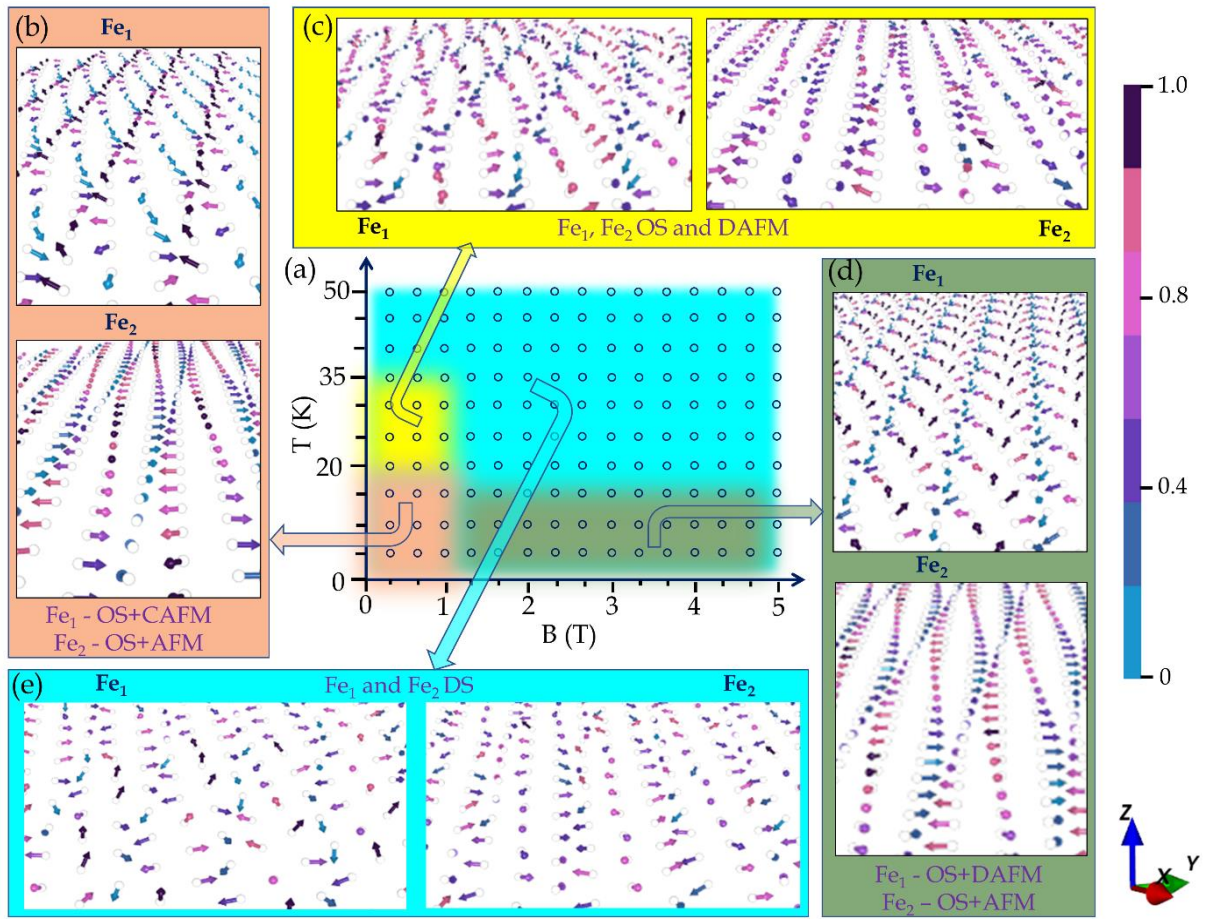

FIG 15. (a) Phase diagram of F4GT. (b-e) The ground-state spin textures of $Fe_1$ and $Fe_2$ for different regions of the phase diagram. The associated color bar describes the *z*-axis projections of the normalized spin moment.

To investigate the temperature and field-dependence of the magnetic long range orders and their respective textures, we have explored the magnetic phase diagram of the system with the applied field values from 0-5 Tesla (T) in the temperature range 0-50 K, as presented in Fig. 15. The *z*-directional magnetic stacking in an unit cell of F4GT contains the $Fe_1$-$Fe_2$-$Fe_2$-$Fe_1$ planar sequence. Each of these planes are seen to have a complex superposition of spin-spirals of both commensurate and incommensurate nature and an AFM order in between the spirals. In general, the $Fe_2$-planes, being encapsulated within the $Fe_1$-planes, are observed to possess more planer orientations of the spins. In the calculated thermomagnetic phase diagram of Fig. 15, we identify four different magnetic phases. The phase boundaries separating these phases are not sharp and have an overlapping nature. The different magnetic phases can be listed as:

(1) around 0-1T and 0-20K, where, both $Fe_1$ and $Fe_2$ planes display an ordered spin-spiral (OS) arrangement along [110]. For $Fe_1$, this OS is superposed with a canted AFM (CAFM) order between the two adjacent spirals. For $Fe_2$, there is a superposition of planar spirals with AFM interaction between the two consecutive spirals. This region is marked with peach color and the corresponding planar magnetic textures are plotted in Fig. 15(b);

(2) from 0-1T and 20-35 K, both $Fe_1$ and $Fe_2$-planes are seen to have an ordered spiral arrangement superposed with a disordered and incomplete AFM (DAFM) order between two adjacent spin spirals. This region is distinguished with yellow color in the phase diagram and the textures are shown in Fig. 15(c);

(3) from 1-5 T and 0-15 K, while $Fe_1$ has an OS, superposed with a DAFM arrangement, $Fe_2$ sticks to the superposition of OS and AFM arrangement. The color of this region is green and the textures are depicted in Fig. 15(d);

(4) beyond 1 T and 35 K, the AFM order completely disappears and the system remains in a disordered spiral state for both $Fe_1$ and $Fe_2$. This region is marked cyan in color with the respective textures in Fig. 15(e). A wide-field view of the texture in the cyan region is presented in Fig. 16. Between 0 and 1 T, the transformation from the OS to the DS configuration at ~ 35 K is marking the onset from the yellow to the cyan zone. For higher values of the applied field, this transformation happens at ~ 15 K.

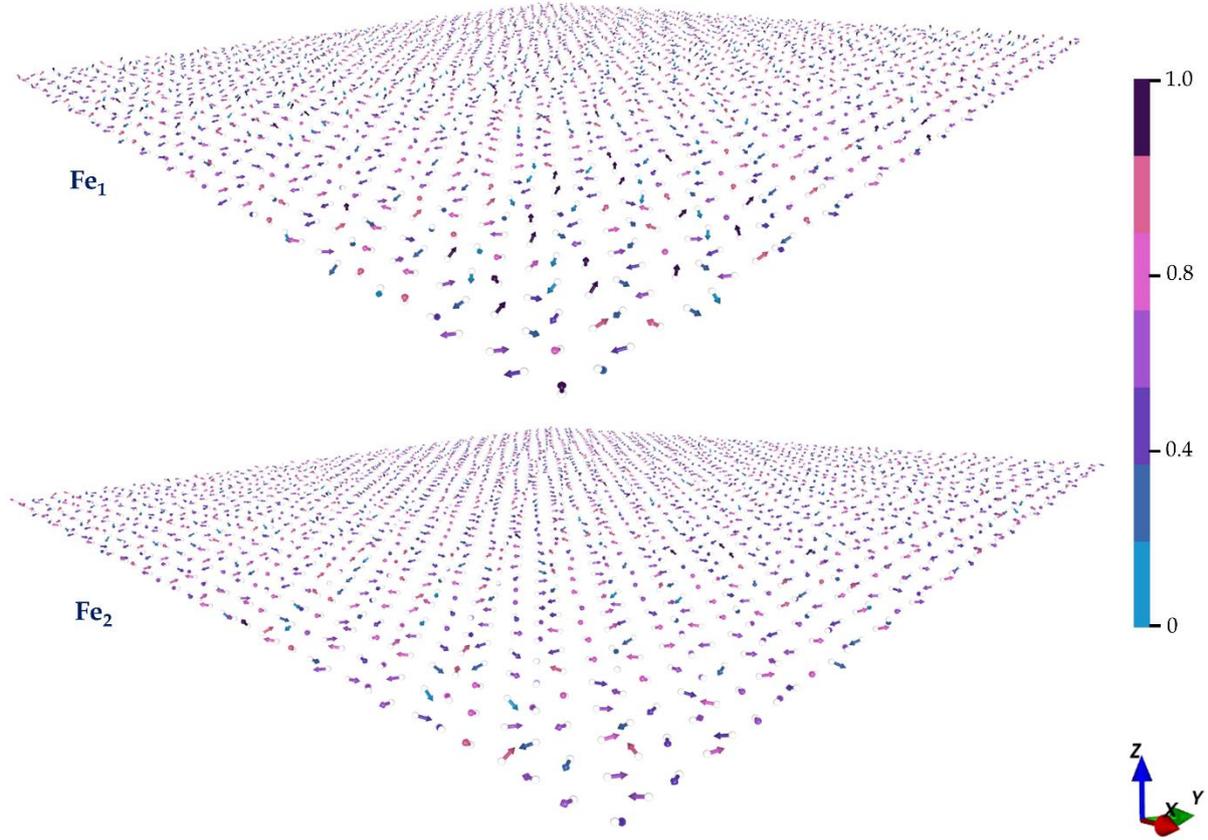

**FIG. 16:** The 3D large-field view of the dynamical ground-state spin-textures at T = 35 K, for each atomic type of $Fe_1$ and $Fe_2$.

In Fig. 17, we have compared the components of the dynamical structure factor plots for 0.01, 30, 35, 40, 78 and 150 K with an applied field of 0.7 T to analyse the impact of this OS-DS transition. Between 30-40 K, which is the experimental crossover region for MR, the out-of-plane (*z*) component of $S(q, \omega)$ has a distinct footprint of the OS-DS transition. In the OS phase (upto 30 K), the distinct features of the planar textures of $Fe_1$ and $Fe_2$ lead to a lifting of degeneracy of the magnon dispersions leading to multiple manon modes for the z-component. From 35 K onwards, the magnon modes become completely degenerate. This feature will be obvious for 35 and 40K spectra in Fig. 17. Moreover, the minima of $S(q,\omega)$ between Γ - T is shifted to the Γ-point for these two temperatures.

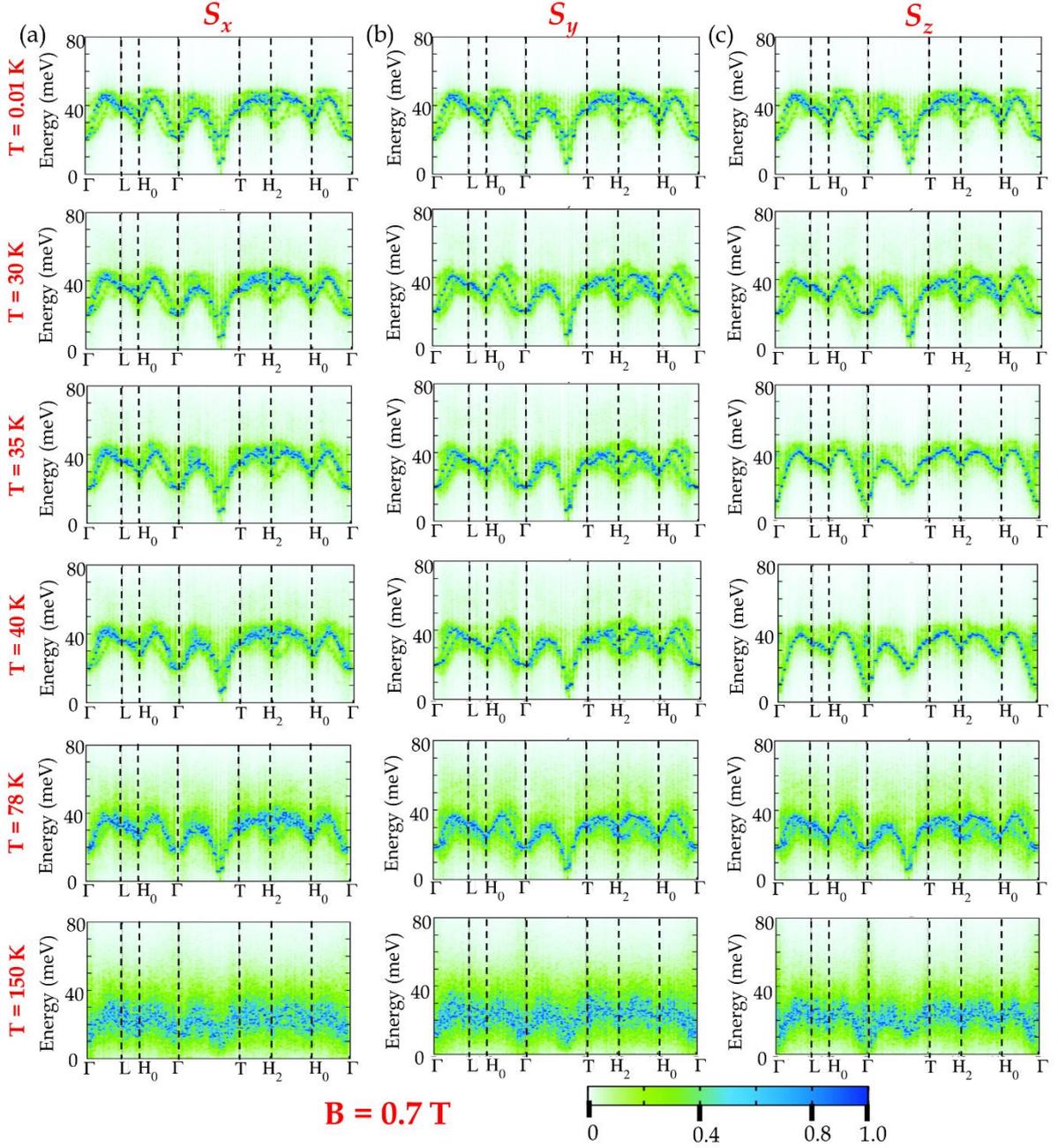

**FIG. 17:** The components of the dynamical structure factors $S(q,\omega)$ plotted along the high-symmetry paths for F4GT for temperatures 0.01, 30, 35, 40, 78 and 150 K with an applied field of 0.7 T.

These two features have their origin in the randomness associated with the spin-orientations due to the OS-DS transition, leading towards a lack of difference between the $Fe_1$ and $Fe_2$ planes. The same transformation could be one of the plausible reasons behind the experimentally observed phase transition at $T_P \sim 38$ K [31], where the measured MR has undergone a crossover from a positive to a negative value, indicating a drop in resistivity. The total resistivity of F4GT will have the impact from the electron-electron, electron-phonon, electron-magnon and magnon-phonon coupling. Whereas the first two contributions will remain unaltered, the OS-DS transformation may impose an impact on the electron-magnon and magnon-phonon coupling. With non-zero applied fields and beyond 35 K, the mode-degeneracy of the magnons may reduce the number of available magnon bands for such coupled scattering. This may affect the electron-magnon or magnon-phonon coupling and lead to a decrease of the measured resistivity for temperatures above 35K.

Thus, our derived spin-dynamical ground states under various thermomagnetic conditions are capable of providing an interpretation of some of the low-temperature phase transitions in the light of a microscopic and atomistic analysis. Moreover, they can also be used as a future reference to plan more detailed experiments on F4GT.

**Conclusion**

We have presented a comprehensive study of the exploration of magnetism in the quasi-2D F4GT compounds with a wide range of first-principles methodologies. Treatment of dynamical electronic correlations with DFT+DMFT+SOC reveals a directional dependence of the field-induced symmetry breaking leading to the opening of a pseudo-gap around Γ-point near $T_{SRT}$ and presence of linearly dispersive band-crossing near $E_F$. The 2D planar FS and its spin-polarization map also corroborates with this observation. The x-ray spectral properties extracted with DFT+DMFT+MLFT have a good agreement with the experimental data. The ab-initio extracted intersite exchange interactions indicate an RKKY-like oscillatory behaviour for all pairwise interactions between $Fe_1$ and $Fe_2$ as well as the presence of significant DM interactions. The calculated phase diagram and the corresponding dynamical magnetic textures show four different regions bearing substantial differences of the underlying long range orders for the two inequivalent sites of Fe. The underlying complex magnetic spin textures extracted by taking into account the dynamical properties of the spin explains the experimental observations quite successfully and, at the same time, provide precious guidelines for future, more complex analyses.

## Acknowledgments

Financial support from Vetenskapsrådet (grant numbers VR 2016-05980 and VR 2019-05304), and the Knut and Alice Wallenberg Foundation (grant numbers 2018.0060, 2021.0246, 2022.0108 and 2022.0079) is acknowledged. O.E. and A.D. also acknowledge support from the Wallenberg Initiative Materials Science (WISE), funded by the Knut and Alice Wallenberg Foundation, for support. We also acknowledge discussions with Atindranath Pal, Riju Pal, Vladislav Borisov, Oscar Grånäs and Biplab Sanyal. DK and MNH acknowledge Kyung-Tae Ko, Korea Basic Science Institute, Daejeon 34133, Korea for providing the .cif file of the F4GT bulk structure. The computations were enabled by both the BARC supercomputing facility and the resources provided by the National Academic Infrastructure for Supercomputing in Sweden (NAISS) and the Swedish National Infrastructure for Computing (SNIC) at NSC and PDC, partially funded by the Swedish Research Council through grant agreement no. 2022-06725 and no. 2018-05973. OE also acknowledges support from STandUPP, the ERC (FASTCORR project) and eSSENCE. This research is part of the project No. 2022/45/P/ST3/04247 co-funded by the National Science Centre and the European Union's Horizon 2020 research and innovation programme under the Marie Skłodowska-Curie grant agreement no. 945339. For the purpose of Open Access, the author has applied a CC-BY public copyright licence to any Author Accepted Manuscript (AAM) version arising from this submission.